\documentclass[useAMS,usenatbib,asm]{mn2e}
\usepackage{graphicx,fleqn,times,color}
\newcommand{\tal}{\it et al. \rm}

\title{Rings and spirals in barred galaxies. III. Further comparisons
  and links to observations.}
\author[Athanassoula \tal]{E. Athanassoula$^1$, M. Romero-G\'omez$^{1,2}$, A. Bosma$^1$, J.J. Masdemont$^3$ \\ 
$^1$Laboratoire d'Astrophysique de Marseille (LAM), UMR6110, 
CNRS/Universit\'e de Provence,\\
Technop\^ole de Marseille Etoile, 38 rue Fr\'ed\'eric Joliot Curie,
13388 Marseille C\'edex 20, France\\
$^2$Institut de Ci\`encies del Cosmos, IEEC-UB, Universitat de Barcelona, Mart\'{i} Franqu\`es 1, 08028 Barcelona, Spain\\
$^3$IEEC \& Dep. Mat. Aplicada I, Universitat Polit\`ecnica de
Catalunya, Diagonal 647, 08028 Barcelona, Spain\\
}
\date{Received }

\begin{document}

\maketitle

\begin{abstract}
In a series of papers, we propose a
theory to explain the formation and properties of rings and spirals in
barred galaxies. The building blocks of these structures are orbits
guided by the manifolds emanating from the unstable Lagrangian points
located near the ends of the bar. In this paper, the last of the
series, we present more 
comparisons of our theoretical results to observations and also give
new predictions for further comparisons. Our theory provides the right
building blocks for the rectangular-like bar outline and for ansae. We
consider how our results can be used to give estimates for the
pattern speed values, as well as their effect on abundance gradients in
barred galaxies. We present the kinematics along the manifold
loci, to allow comparisons with the observed kinematics along 
the ring and spiral loci. We consider gaseous arms and
their relations to 
stellar ones. We discuss several theoretical aspects and
stress that the orbits that constitute the building blocks of the
spirals and rings are chaotic. They are, nevertheless, spatially 
well confined by the manifolds and are thus able to outline the relevant
structures. Such chaos can be termed `confined chaos' and can play a
very important role in understanding the formation and evolution of
galaxy structures and in galactic dynamics in general. This work, in
agreement with several others, argues convincingly that galactic
dynamic studies should not be limited to the study of regular motions
and orbits.   
\end{abstract}


\section{Introduction}
\label{sec:intro}

In previous papers (\citealt{RomeroGMAG06}, hereafter Paper I;
\citealt{RomeroGAMG07}, hereafter Paper II; \citealt{AthaRGM09},
hereafter paper III; \citealt{RomeroGMGA09}) we proposed a theory to
explain the formation and structure of spirals and inner and outer rings
in barred galaxy potentials. We propose that the building blocks of these
structures are chaotic orbits that are guided by manifolds associated
with the Lagrangian points $L_1$ and $L_2$ which are located along the
direction of the bar major axis. These manifolds can be thought of as
tubes that confine the orbits, so that the latter
can form thin structures in configuration space.   

A theory, however, can be dynamically correct but still
irrelevant to a particular application. We therefore have to check whether
our theory is applicable to spirals and rings observed in barred disc
galaxies. It is thus necessary to compare our theoretical results and
predictions to observations. We started this in Paper IV
(\citealt{AthaRGBM09}, hereafter Paper IV) where we made a number of
comparisons of model spirals and rings, the latter both inner and outer, to
observations and found good agreement. We found that our theory can
reproduce all necessary morphologies of both inner and outer rings,
and produces no unrealistic morphologies. Model inner rings were found
to be elongated along 
the bar and outer ones either along it ($R_2$, or $R_2'$), or
perpendicular to it ($R_1$, or $R_1'$), or both ($R_1R_2$). 

Model spirals in barred galaxy
potentials were found to be 
predominantly two-armed and trailing, although arms of higher
multiplicity are possible for specific potentials. They were found to
have the right shape and could reproduce the fall-back of an arm
towards the bar region or towards the other arm, which is observed in
many spirals. A quantitative comparison to the arm shapes of NGC 1365
proved successful. We predict that the relative strength of the
non-axisymmetric forcing in the region around and somewhat beyond
corotation influences the winding of the arms, in the sense that in
strongly barred galaxies the spirals will be more open than in less
strongly barred ones. Thus, a series of models with increasing bar
strength will have a continuous sequence of morphologies from $R_1$ to
$R_1'$, then to tightly wound 
spirals, to end with open spirals. We also compared the shape of the inner and
outer rings, as well as the ratio of outer-to-inner ring major axes of our
models to observations and discussed which type of potentials give
best agreement. We also showed that there are correlations, or trends,
between all the ratios of ring sizes discussed above and the bar
strength. These correlations are very tight if
only one type of model potential is used but thicken, sometimes very
considerably, if other models, with different properties and
analytical expressions, are included. 

The present paper is the fifth and last of this series. A 
reminder of the main theoretical results and prerequisites is given in
Sect.~\ref{sec:theory}. This is very brief, since these concepts have
been described extensively in Papers I, II and III.
Here we will mainly present further comparisons with   
observations and predictions. These include e.g. the bar shape
(Sect.~\ref{subsec:barshape}), ansae (Sect.~\ref{subsec:ansae}),
radial drift and abundance gradients 
(Sect.~\ref{subsec:abundancegr}). Kinematics and behaviour of the
line-of-sight velocities are discussed in Sect.~\ref{sec:Kinematics}
and pattern speed prediction is the subject of Sect.~\ref{sec:omegap}. 
In Sect.~\ref{sec:discussion} we discuss specific
topics, such as time evolution, self-consistency and spiral structure
formation, and we also compare with other spiral structure theories
and give outlines for further work. These discussions rely on material
from all five papers in this series. Finally, we
summarise and conclude in Sect.~\ref{sec:summary}. 
 
\section{Theoretical reminders}
\label{sec:theory}

Our theory relies to a large extent on the dynamics of the Lagrangian points
$L_1$ and $L_2$ of a two-dimensional barred galaxy system. These
points are
located along the direction of the bar major axis and are, in the
standard case, saddle point unstable \citep{Binney.Tremaine08}. Their
distance from the centre is denoted here by $r_L$ and it is referred to
as corotation, or Lagrangian radius. Each of them is surrounded by a family of 
periodic orbits, called Lyapunov orbits \citep{Lyapunov49}. Since
these are unstable they can not trap around them quasi-periodic orbits
of the same energy\footnote{We will all through this paper loosely
  call `energy' the numerical value of the Hamiltonian in a frame of
  reference co-rotating with the bar, i.e. a
  frame in which the bar is at rest, and denote it by $E_J$.}, so that
any orbit in their 
immediate vicinity (in phase space) will have to escape the
neighbourhood of the corresponding Lagrangian point. Not all departure
directions are, however, possible. The direction in which the orbit
escapes is set by what we call the invariant manifolds. These can be
thought of as tubes that guide the motion of particles of the same
energy as the manifolds \citep{GomezKLMMR, KoonLMR}. Each manifold has
four branches emanating 
from the corresponding Lyapunov orbit, two of them inside corotation
(inner branches) and two of 
them outside (outer branches). Along two of these branches (one inner and one
outer) the mean motion is towards the region of the Lagrangian point
(stable manifolds), 
while along the other two it is away from it (unstable manifolds). We
need to stress that the terms `stable' and `unstable' do not
mean that the orbits that follow them are stable and unstable,
respectively. In fact all the orbits that follow the manifolds are
chaotic, but they are in a loose way `confined' by the manifolds, so
that they stay together, at least for a few bar rotations, in what
could be called a bundle. 
These manifolds do not exist for all values of the energy, but only for
energies for which the corresponding Lyapunov periodic orbit is
unstable. This means energies within a range starting from the energy
of the $L_1$ or $L_2$ ($E_{J,L_1}$) and extending over a region whose extent
depends on the model \citep{SkokosPA02}. In this series of papers we
propose that these manifolds and orbits 
are the building blocks of the spirals and rings in barred galaxies.  

As shown in Papers II and III, the properties of the manifolds depend
strongly on the relative strength of the non-axisymmetric forcing at
and somewhat beyond corotation. We measure this with the help of the
quantity $Q_{t,L_1}$, which is the value of $Q_t$

\begin{equation}
Q_t (r) = (\partial \Phi (r, \theta) /\partial \theta)_{max}/(r\partial
\Phi_0/\partial r),
\label{eq:Qt}
\end{equation}

\noindent
at $r_L$, the radius of the Lagrangian point $L_1$ or $L_2$, i.e. 
$Q_{t,L_1}=Q_t (r=r_L)$. This is not the same as
the strength of the bar, which is often measured by $Q_b$, i.e. the maximum
of $Q_t$ over all radii shorter than the bar extent. 
The radius at which this maximum occurs can be small compared
to $r_L$. So $Q_b$ is not necessarily a good proxy for 
$Q_{t,L1}$ and vice versa. Nevertheless, for the sake of brevity,
we will often replace in our discussions ``non-axisymmetric
forcings which are relatively strong at and beyond corotation'' simply
by ``strong bars'', or ``strong non-axisymmetric forcings''. 

The models used here are described extensively in an appendix of
Paper III. Model A  
  is taken from \cite{Atha92a} and has a Ferrers bar \citep{Ferrers77}
of semi-major axis $a$, axial ratio $a/b$, quadrupole moment $Q_m$ and
rotating with a pattern speed $\Omega_p$. The central concentration of
this model is characterised by its central density $\rho_c$. The
fiducial model of this series of papers has the parameter values $Q_m
= 4.5\times 10^4$, $a/b$ 
= 2.5, $r_L$ = 6, $a$ = 5, $\rho_c = 2.4\times 10^4$ and $n$ = 1. The
units are $10^6 M_{\odot}$,  1 kpc and 1 km/sec, for the mass, length 
and velocity, respectively. Model D
has a Dehnen-like bar potential \citep{Dehnen00} characterised by a
strength parameter ($\epsilon$) and a scale length ($\alpha$). The BW
model has a bar with a Barbanis \& Woltjer type potential
\citep{Barbanis.Woltjer67} characterised by a strength parameter
($\hat\epsilon$) and a scale length ($r_1$). As in the previous
papers, we use the two latter models to represent forcings not only from
bars, but also from spirals, oval discs and triaxial haloes.

Unless otherwise stated, we use in this series of papers a frame of
reference corotating with the bar, because calculations are more
straightforward in this frame. We measure angles in the standard 
mathematical sense, i.e. starting from the positive $x$ axis and
increasing anticlockwise.

\section{Further comparisons with observations}
\label{sec:otherprop}

\subsection{Bar shape}
\label{subsec:barshape}

\begin{figure} 
\centering
\includegraphics[scale=0.3,angle=-90.0]{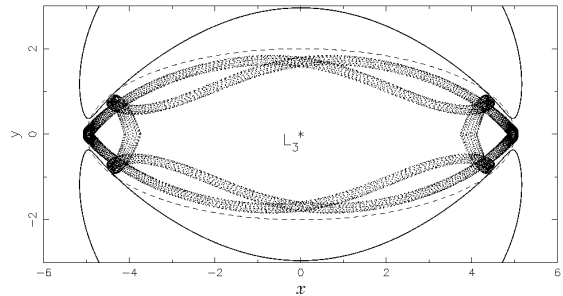} \hspace{0.5cm}
\caption{Shapes outlined by the inner manifolds after half a
  revolution around the galactic centre for a model with strongly unstable $L_1$
  and $L_2$ Lagrangian points. We see that these provide building
  blocks for a rectangular-like outline of the bar. The dashed line
  traces the outline of the Ferrers potential and the solid lines the
  curves of zero velocity for the same energy as the manifolds. The
  centre of the galaxy, where the $L_3$ Lagrangian point is located,
  is marked with a star.} 
\label{fig:barshape-stand}
\end{figure}

\begin{figure} 
\centering
\includegraphics[scale=0.3,angle=-90.0]{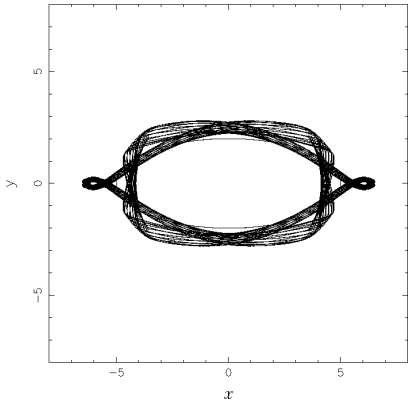} \hspace{0.5cm}
\caption{The inner manifolds for a model with stable $L_1$
  and $L_2$ Lagrangian points (see Paper III). It is clear that these
  provide building blocks for a rectangular-like outline of the
  bar, as well as for ansae.} 
\label{fig:barshape-stable}
\end{figure}

\citet{AthaMWPPLB90} showed that the isodensities of the outer bar
region of early type, strongly barred galaxies have predominantly a
rectangular-like shape. This was later confirmed for larger samples 
by \citet{Gadotti08, Gadotti10} and for bars in $N$-body simulations by
\citet{Atha.Misiriotis02}. The origin of this shape has been
widely debated. A first obvious possibility, already discussed by
\citet{AthaMWPPLB90}, is that the rectangularity is due to orbits trapped
around the periodic orbits in the vicinity of inner ultraharmonic
(4:1) resonance. This, however, could only be true if the
corresponding periodic orbits are stable, which  is not generally true
\citep{Contopoulos88,Atha91,Atha96,PatsisAQ97}. \citet{Atha91,Atha96}
proposed 
alternative solutions, like trapped regular orbits wobbling around
x$_1$ periodic orbits, superposition of orbits trapped around two families of
3:1 periodic orbits symmetric with respect to the bar minor or major axis, or
chaotic orbits confined by cantori, i.e. orbits staying for long times around
unstable rectangular-like 4:1 orbits. All these alternatives were
examined in detail in the potential of NGC 4314 \citep{PatsisAQ97}. 

Our work here argues that the role of chaotic orbits in explaining
this rectangular-like shape is crucial. Indeed, in Paper III we
discussed the inner branches of the manifolds and showed that their
properties depend on the bar strength. For weak bars, if 
we trace the inner manifolds over more than half a revolution around the
galactic centre, then the manifolds
retrace the same loci, forming an inner ring. On the other hand, for
strong bars, the inner
manifolds cover a new path after the first revolution around the
galactic centre and retrace the same loci only after a couple or a few
revolutions. An example of the latter is shown in
Fig.~\ref{fig:barshape-stand}. The bar here is strong ($Q_m$ = 7 and 
$r_L$ = 5, the values of the remaining parameters being the same as for the
fiducial model) and it is clear that the orbits
corresponding to such manifolds can provide building blocks for the
rectangular outline of the bar.   
Fig.~\ref{fig:barshape-stable} shows another example, this time from a
case with stable $L_1$ and $L_2$ Lagrangian points (Paper III). In
this case also, 
the inner manifold branches can provide useful building blocks for the
rectangular bar outline.

In both these examples, and in all the other models we tried,
considerable parts  of the inner manifold branches can
outline the outer part of the bar so that this rectangularity of the
isodensities should be expected to occur in the outer parts of the
bar. This is indeed what is seen both from the observations
(\citealt{AthaMWPPLB90}) and the simulations
(\citealt{Atha.Misiriotis02}). As discussed in Paper III, such building
blocks for rectangular-like bar outlines should be
seen mainly in strong bars and this again is in good agreement with
both observations (\citealt{AthaMWPPLB90,Gadotti08,Gadotti10}) and 
simulations (\citealt{Atha.Misiriotis02}). 

To summarise, the inner manifold branches can produce the right building
blocks for the rectangular-like outline of the outer parts of
strong bars. 

\subsection{Ansae}
\label{subsec:ansae}

Ansae are concentrations of matter located at both ends of the bar
\citep{Sandage61, Atha84, MartinezVKB07}. Manifolds can, in some cases,
contribute the right building blocks for this structure as
well. In general, the $L_1$ and $L_2$ are unstable. In Paper III,
however, we showed that if there are mass concentrations centred on
these unstable Lagrangian points, then the $L_1$ and $L_2$ can become
stable, provided these masses are sufficiently massive and
concentrated. In this case, four other unstable points will appear on the
direction of the bar major axis, one on each side of the $L_1$ and of
the $L_2$. There are thus seven rather than
three Lagrangian points on the $x$ axis (i.e. the direction of the bar
major axis). The central one ($L_3$), two at smaller radii than the
$L_1$ and $L_2$ (which we call $L_1^i$ and $L_2^i$, where $i$ stands
for inner) and two at larger radii (which we call $L_1^o$ and $L_2^o$,
where $o$ stands for outer). In such configurations, part of the
manifold, emanating from $L_1^i$ 
or $L_1^o$ (respectively $L_2^i$ or $L_2^o$), encircles the $L_1$ (or
$L_2$). This part of the manifolds, together with 
regular orbits trapped around the now stable Lyapunov orbits 
can be the building blocks for ansae (Fig.~\ref{fig:barshape-stable}). 

%Thus, similarly
%to the inner manifolds of the standard case, such inner
%manifolds could contribute to the rectangular like isodensities of
%strong bars in early type barred galaxies \citep{AthaMWPPLB90}, as
%will be discussed in Paper IV. The
%second point is that the outermost parts of this structure, composed
%by material between $L^i_1$ and $L^o_1$ ($L^i_2$ and $L^o_2$), has the
%characteristic shape and location of a structure often observed in
%early type barred galaxies and called ansae \citep{Sandage61, Ath84,
%  MartinezVKB07}. Ansae are mass concentrations, lying along the bar
%major axis and just after the end of the bar. The part of the
%manifolds between $L^i_1$ and $L^o_1$ ($L^i_2$ and $L^o_2$), together
%with orbits trapped around the stable orbits surrounding $L_1$
%($L_2$), could be the building blocks of ansae.  

\subsection{Motions within the galaxy and abundance gradients}
\label{subsec:abundancegr}

\begin{figure*}
\begin{center}
\includegraphics[scale=0.8,angle=-90.]{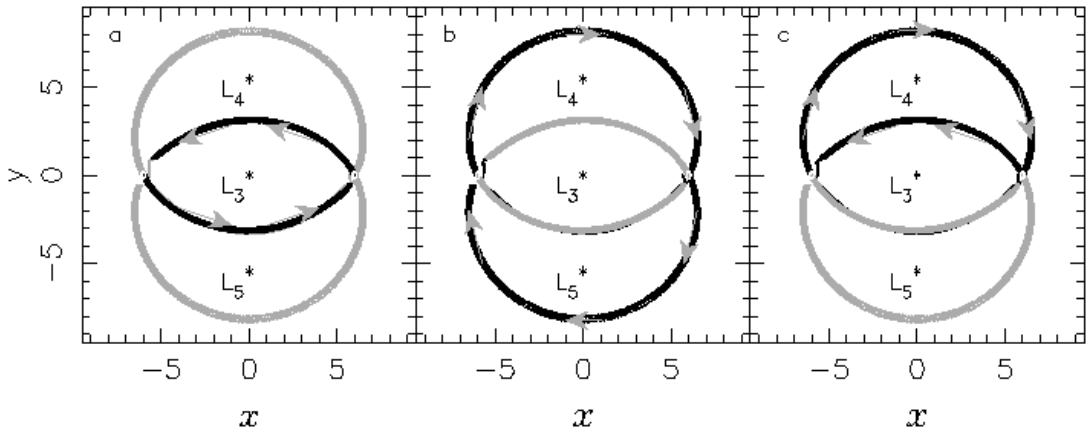}
\end{center}
\caption{Three possible circulation patterns for particles 
on orbits guided by the manifolds in an example with an $rR_1$
morphology. Any given such particle can stay on 
the inner branches of the manifold, i.e. can trace the inner ring
(left panel), or it can stay on the outer branches of the manifold,
i.e. can trace the outer ring  (middle panel). Alternatively, it can
trace a path that includes one inner and one outer branch. Two such
paths are possible. One is shown in the right panel. The other, i.e. the fourth 
alternative, is not plotted here since it is similar to the third one, except 
that it is symmetric with respect to the bar major axis. After a
particle has completed a full circulation path, it can either repeat
the same path, or take any of the other three ones. The manifolds
coming into play in each case are plotted in black in the
corresponding panel, and the remaining ones in light grey. In white,
we plot the Lyapunov periodic orbits of the corresponding
energy level. The arrows show the direction of the motion. This
particular model has $Q_m$ = 8 and $r_L$ = 6, the remaining parameters
being the same as for the fiducial model. Similar results, however,
are found for all $rR_1$ morphologies. 
}
\label{fig:circring}
\end{figure*}

As already discussed in Papers I to III, the $L_1$ and $L_2$
Lagrangian points can be considered as gateways through which material
from within corotation can move to larger radii outside corotation and
vice versa. In general, this 
will induce considerable radial motion within the galaxy
and will therefore partly smooth out abundance gradients. We will discuss here
in more detail two specific cases, an $rR_1$ morphology and a
spiral morphology. As usual, we place ourselves in a frame of
reference rotating with the bar. In this frame of reference the sense
of rotation in the outer ring, be it $R_1$ or $R_2$, 
is clockwise, i.e. retrograde.
On the other hand, it is anticlockwise, i.e. direct in this frame,
for the inner manifold branches, i.e. for the inner ring, or the outer
parts of the bar.

A star trapped by manifolds resulting in an $rR_1$ morphology can follow four
different paths. In the first one, illustrated in the left panel of
Fig.~\ref{fig:circring}, the star will be guided solely by the inner
manifold branches. It will thus trace the inner ring, or the outer
part of the bar, without moving to regions further out. Such motions
have been found in many cases, both for models and observations, as 
e.g. in the hydrodynamic models of IC 4214 by
\cite{Salo.RBPCCL99}. They will only smooth out abundance
gradients within the outer parts of the bar.

In the second alternative, illustrated in the central panel of
Fig.~\ref{fig:circring},  the star will be guided solely by the outer
manifold branches. If most stars followed such paths, there would be some
smoothing of the abundance gradients because the paths are far from circular,
but this smoothing would be considerably less than what will occur for
the paths described below, since it concerns only an annulus outside
corotation.  

The third path, illustrated in the right panel of 
Fig.~\ref{fig:circring}, is more complex. A mass element near $L_1$
will move on the inner branch away from $L_1$ and towards $L_2$. When
it reaches the vicinity of $L_2$ it will leave it, guided by the
unstable outer manifold branch. It will thus move outwards until it
reaches a maximum radius and then continue inwards towards  $L_1$ on
the stable outer manifold branch to reach again the vicinity of $L_1$
from which it started. The sense of the motion is clockwise with
respect to the Lagrangian point $L_4$, which is enclosed within
it. The fourth alternative is similar to the third 
one. Its loci are the mirror image of those of the third one with
respect to the bar major axis and the sense of rotation is
clockwise with respect to $L_5$. For the last two paths, the radial
mixing of material concerns a region which is radially much more
extended than for the first two and could thus smooth out
considerably any radial abundance gradients. Even so, this will not
concern material at radii smaller than the minor axis of the inner
ring or larger than the major axis of the outer ring. Once a star has
completed one path it can continue on the same path following again
the same circulation pattern, or it can take any of the other three
paths. Thus a star can first go around the inner ring, then around the
outer ring and then follow a mixed trajectory, or any other sequence
of the four possible circulation patterns.

\begin{figure}
\begin{center}
\includegraphics[scale=0.45,angle=-90.]{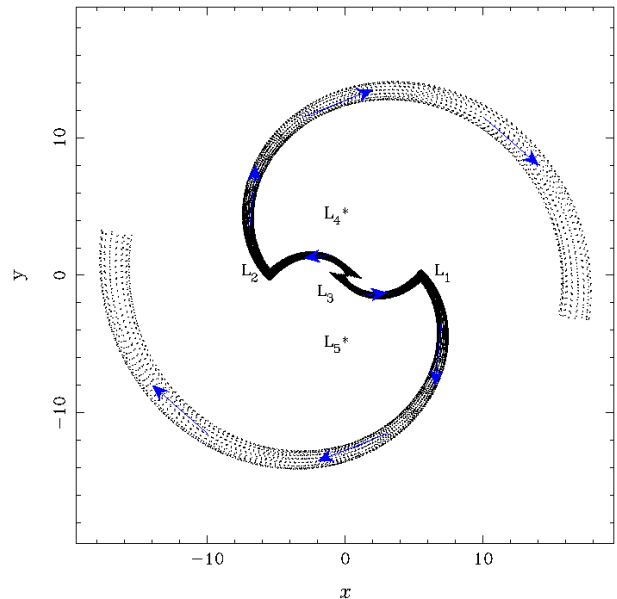}
\end{center}
\caption{Circulation patterns for particles in a model with spiral
  morphology. Such patterns induce very strong radial mixing, since
  material from very near the galactic centre can reach regions far
  out in the disc. 
}
\label{fig:circsp}
\end{figure}

If instead of an $rR_1$ we have a spiral morphology, then the
circulation pattern is much simpler and the amount of radial mixing
yet more important. Material moves along one of the inner stable
manifold branches towards one of the unstable Lagrangian points $L_1$
or $L_2$. Since
for spirals the non-axisymmetric forcing is strong, the inner
manifolds traverse regions of the bar located at a relatively small
distances from the centre and thus can guide bar material from these
regions outwards (see Fig.~\ref{fig:circsp}).
Such material will then  
continue outwards along the spiral, so that, starting from well within
the bar, it will reach radii very far from the centre. In such cases,
the radial mixing concerns a very large radial extent and this
outgoing material can, furthermore, extend the disc outwards. This
motion should also reduce considerably any 
abundance gradients in these regions. It is, however, not possible to
make more precise predictions 
about the abundance gradients, unless a full model is built,
including the abundance gradients before the bar was introduced.  

\cite{Martin.Roy94} studied abundance gradients in disc galaxies and
found that these are shallower in barred than in non-barred ones. 
They furthermore found that the strength of the bar plays a
determining role: the flatter gradients are found in galaxies with
stronger bars. These results are in very good agreement with our
theory. Indeed, circulation of material driven by the
manifolds will render abundance gradients shallower. For bar strengths
below a given threshold the morphology is $rR_1$ (Paper III) and, as
we saw above, in such cases the region in which the radial mixing
occurs is set by the elongation of the rings. Furthermore, both
observations and theory (Paper IV), show that rings are more
elongated when the bars are stronger. Therefore, in more strongly
barred galaxies the radial mixing will occur over regions which are more
radially extended and thus lead to flatter abundance gradients. For
yet stronger bars the manifolds 
will drive a spiral morphology, in which, as discussed above, the
radial mixing will concern yet larger regions. Thus our theoretical
results are in good agreement with observations on this point. 

\section{Kinematics}
\label{sec:Kinematics}

Kinematics can provide very strong constraints in comparisons between
models and observations. For this reason, we discuss in this section
the velocities along the orbits guided by the manifolds in two
specific morphologies, the $rR_1$ rings and the spirals, which are the most
frequently observed cases \citep{Buta95} and are also the most
interesting dynamically. Since our aim here is to provide
measurements that can be compared with observations, we will calculate
the velocities in the inertial frame of reference. 

\begin{figure} 
\centering
\includegraphics[scale=0.44,angle=-90.0]{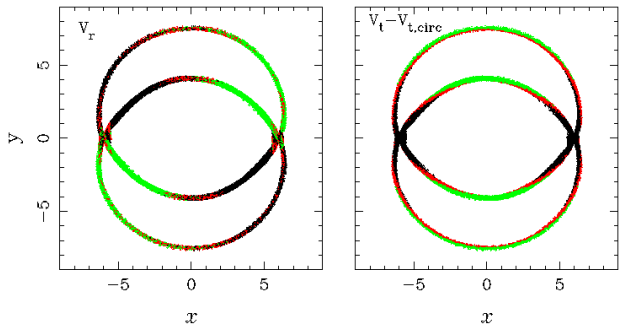} 
\caption{Radial (left panel) and tangential velocities (after
  subtraction of the circular velocity -- right panel) along the inner
  and outer rings of an $rR_1$ morphology case. The velocities are
  calculated in an inertial frame of reference. Positive values are given
  in black and negative in green. Values around zero, more precisely in
  the (-2.5, 2.5) bracket, are given in red. 
}
\label{fig:rR-vrvt-xy}
\end{figure}

\begin{figure} 
\centering
\hspace{0.75cm} {\bf Inner ring}
\includegraphics[scale=0.44,angle=-90.0]{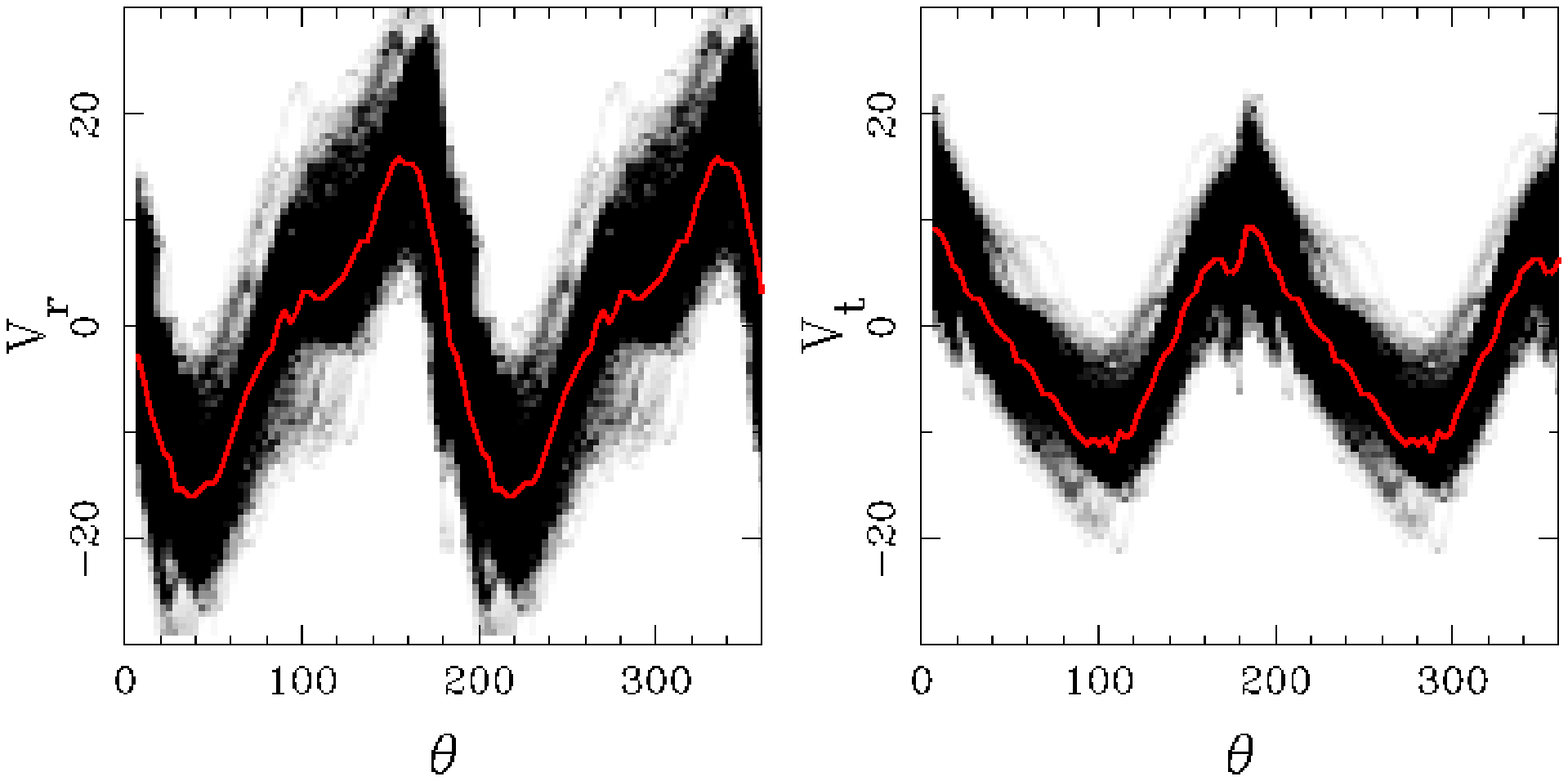} \\
\vspace{0.5cm}
\hspace{0.75cm} {\bf Outer ring}
\includegraphics[scale=0.44,angle=-90.0]{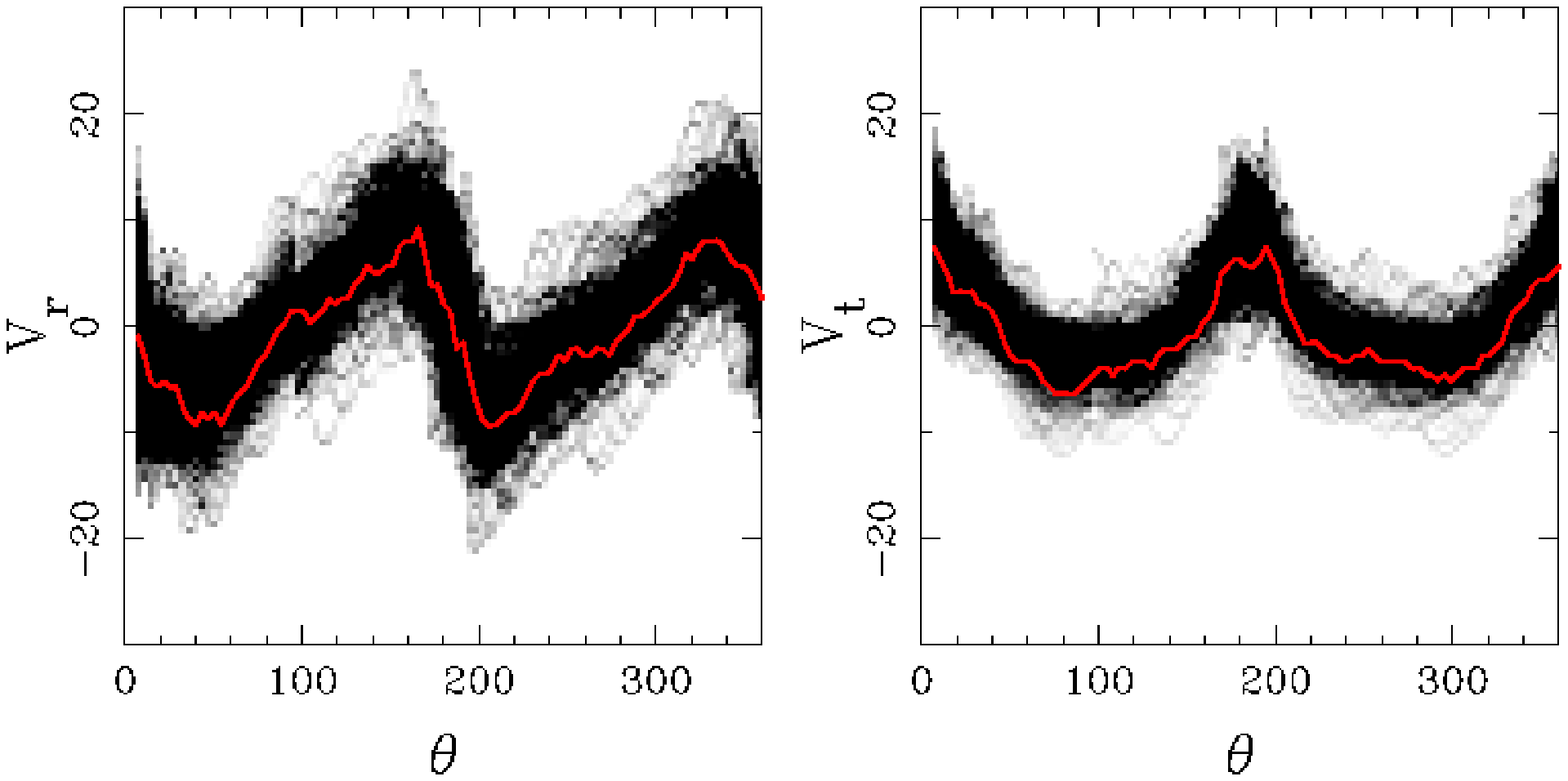} 
\caption{Radial (left panels) and tangential velocities (after
  subtraction of the circular velocity -- right panels)
  along the inner (upper panels) and outer (lower panels) rings. The
  model is the same as in Fig.~\ref{fig:rR-vrvt-xy}. The median values
  as a function of the angle are given by the red curve. 
}
\label{fig:rR1-vrvt-theta}
\end{figure}

\begin{figure*} 
\centering
\hspace{0.8cm}{\bf Inner ring}\\
\includegraphics[scale=0.65,angle=-90.0]{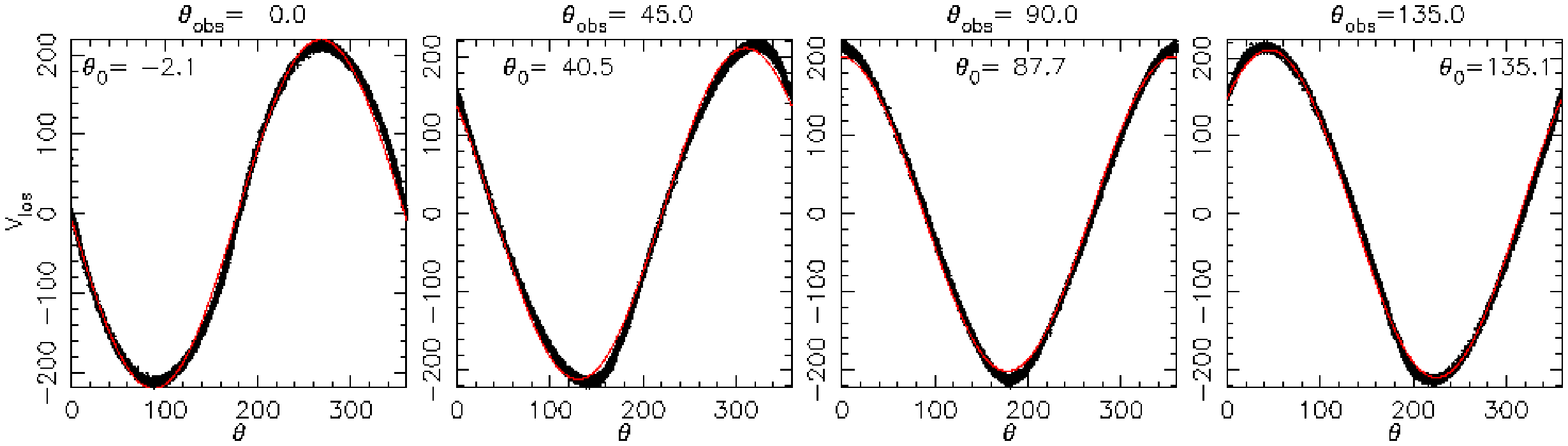} \\
\vspace{0.75cm}
\hspace{0.8cm}{\bf Outer ring}\\
\includegraphics[scale=0.65,angle=-90.0]{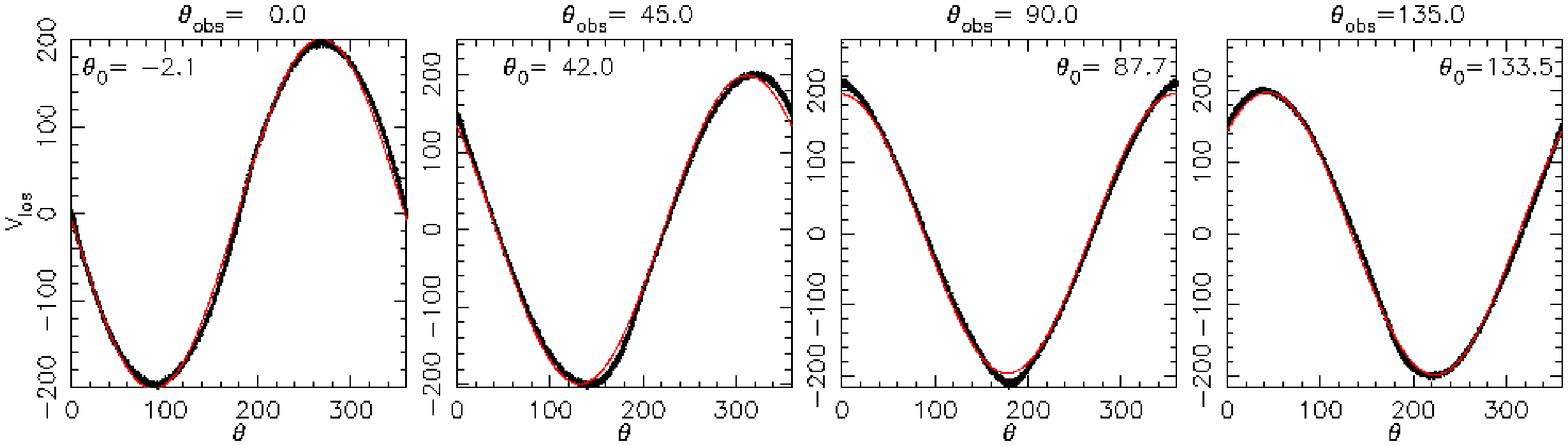} 
\caption{Line-of-sight velocity for points along the inner (upper row of
  panels) and outer (lower row of panels) ring in the case of an
  $rR_1$ morphology. Each panel corresponds
  to a given viewing angle, whose value is given at the top of the
  panel. The
  models and orbits are as in Figs.~\ref{fig:rR-vrvt-xy} and
  \ref{fig:rR1-vrvt-theta}. The red lines are the best fitting
  sinusoidal curves. All angles are in degrees.
}
\label{fig:sinfitrR}
\end{figure*}

Fig.~\ref{fig:rR-vrvt-xy} gives information on the motion along the
manifolds in the case of an $rR_1$ morphology\footnote{Note
  that this is different from what is given in
  \cite{Melnick.Rautiainen09}, where the velocities are measured in a
  circular annulus, only part of which corresponds with the ring. Their
  results, therefore, are not directly comparable to
  ours.}. 
The model here, as well as in Figs. \ref{fig:rR1-vrvt-theta} and
\ref{fig:sinfitrR}, is the fiducial A model. It has $Q_{t,L_1}$ =
0.038 
and its inner and outer rings have axial ratios of 0.81 and 0.90,
respectively. However, results from other models  
with a similar morphology are qualitatively the same. As expected from
the results presented so far and as can be seen in the left panel of
Fig.~\ref{fig:rR-vrvt-xy}, along the outer ring the motions is 
outwards in the upper left quadrant of the ring and inwards in
the upper right part. Zero radial
velocity is reached roughly at the direction of the major axis of the
outer ring. This is in
agreement with the fact that the motion is outwards (i.e. away from the
Lagrangian point from which the manifold emanates) along the unstable
manifold branches and inwards along the stable ones (Papers I, II and III).  

In order to show best the perturbations of the tangential velocity,
we subtract from it the circular velocity calculated at the radius at
which the measurement is made. The result is given in the right panel
of Fig.~\ref{fig:rR-vrvt-xy}. This shows that, in general, the
tangential motion in the outer part of the ring is slower than in the
inner part. 

Fig.~\ref{fig:rR1-vrvt-theta} shows, for each ring separately, the
radial and tangential components of the velocity, calculated as above
along the ring, but now plotted as a function of the azimuthal
angle. In both cases, both velocity components show an oscillation 
of an amplitude less than or of the order of 20 km/sec. Also, in both
cases the oscillation has two maxima and
two minima, but is not sinusoidal. The radial velocity curve is
asymmetric, in the sense that the angular extent over which
the radial velocity increases with increasing angle is more extended 
than that over which it is decreasing. Some of these results could have been
directly obtained from the manifold analyses done so far (see e.g. 
Fig.~\ref{fig:circring}).
Material leaving the $L_1$ or the $L_2$ Lagrangian point will move away
from it on the unstable manifold branches. Since the major axis of the outer
ring is oriented perpendicular to the bar, this means that the radial
velocity will be positive until roughly 90$^\circ$ for the $L_2$, or
-90$^\circ$ for the $L_1$. After that, the sense of the radial motion
is reversed and the particle will head for the Lagrangian point
opposite to the one it started from. This explains the changes of sign
in the lower left panel of Fig.~\ref{fig:rR1-vrvt-theta}. For the
inner ring, both the sense of circulation (in the rotating frame of
reference) and the directions of the major and minor axes are
reversed, so that the variations of the sign of the velocity as a
function of $\theta$ stay globally the same.
Fig.~\ref{fig:rR1-vrvt-theta} also shows that the 
amplitude of the oscillations is larger for the inner than for the
outer ring, which should be linked to the fact that the bar forcing is
stronger at the inner ring radii than at the outer ring and the inner
ring is
more elongated than the outer one (for observed rings see
\citeauthor{Buta95} \citeyear{Buta95},
and for theoretical ones see Paper IV, Sect. 5).   

The tangential velocities, calculated in the inertial frame of
reference and after subtraction of the circular velocity, are given in
the right panels of Fig.~\ref{fig:rR1-vrvt-theta}. Again, the maxima
and minima occur at roughly the same angles for the inner and the outer
rings and the amplitude of the oscillation is much larger for the
inner than for the outer ring. 
 
To compare with observations we need to calculate the component of the
velocity along the line of sight, $v_{los}$. The results are given in
Fig.~\ref{fig:sinfitrR}, 
separately for the inner and the outer rings and for four viewing
angles. The line-of-sight velocities look roughly sinusoidal, so, as is
customary for observations, we fitted sine curves of the form $A~sin
(\theta - \theta_o)$ to them. The values
of the amplitude $A$ vary between 202 (for $\theta_{obs} = 90^\circ$)
and 219 (for $\theta_{obs} = 0^\circ$ and 180$^\circ$) and the
corresponding $\theta_0$ values indicate the position at which the
line-of-sight velocity is zero. The latter are very near the viewing angle
of the observer and always lagging slightly behind it. The minimum
deviation is less than a degree (for $\theta_{obs}$ = 135) and the
maximum deviation around 4$^\circ$ (for $\theta_{obs}$ = 45). These
values are in good agreement with the differences between the
inner ring photometric and kinematic major axis position angles observed
e.g. for NGC 6300 \citep{Buta87} and NGC 1433 \citep{Buta86}. Similar
results are found for the outer $R_1$ ring (two bottom panels), but
the amplitude varies little with $\theta_{obs}$.  

\begin{figure} 
\centering
\hspace{0.75cm}{\bf Inner ring}\\ 
\includegraphics[scale=0.44,angle=-90.0]{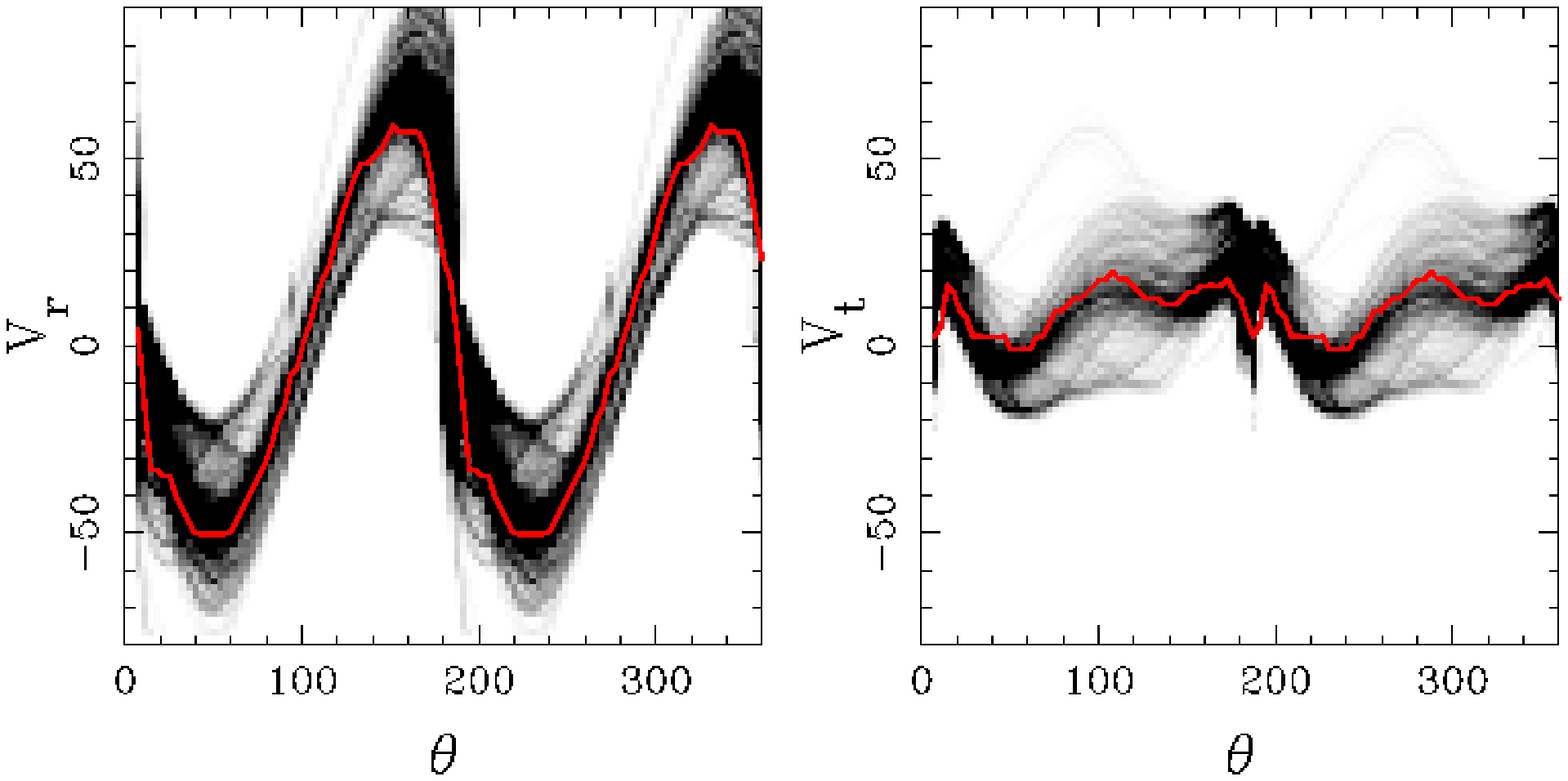} \\
\vspace{0.5cm}
\hspace{0.75cm}{\bf Outer ring}\\
\includegraphics[scale=0.44,angle=-90.0]{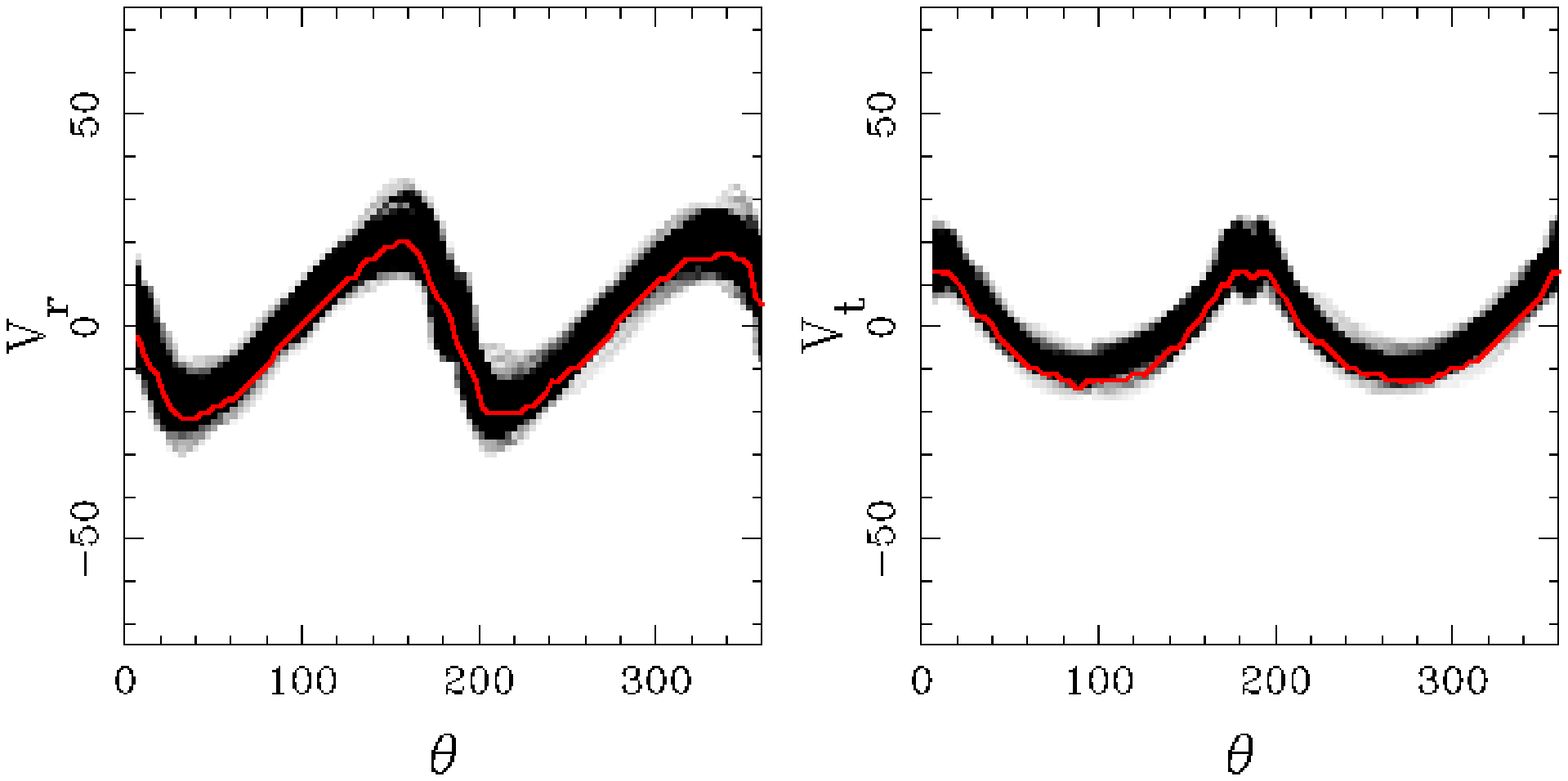} 
\caption{As Fig.~\ref{fig:rR1-vrvt-theta}, for 
  a model with a stronger bar ($Q_m$ = 9 $\times 10^4$), but still
  with an $rR_1$ morphology. The red curve gives the median value
  as a function of the angle.
}
\label{fig:rR1-vrvt-theta-strong}
\end{figure}

Let us now look at the kinematics of a case with a somewhat stronger relative
non-axisymmetric forcing, which, in agreement with what was found in
Paper III, has also an $rR_1$ morphology, but more elongated
rings (minor-to-major axes ratios of 0.57 and 0.78 for the inner
and outer rings, respectively). The model has the same
parameters as the one we analysed 
above, but a $Q_m$ = 9 $\times 10^4$. The radial and tangential
velocities along the ring loci are given in
Fig.~\ref{fig:rR1-vrvt-theta-strong}. The main difference with the
previous case is that the amplitude of the oscillations is much larger,
so that the absolute value of the radial velocity can go even higher
than 50 km/sec at certain angles, while that of the peculiar
tangential velocity can go as high as, or even exceed 25 or 30 km/sec. 

\begin{figure} 
\centering
\includegraphics[scale=0.44,angle=-90.0]{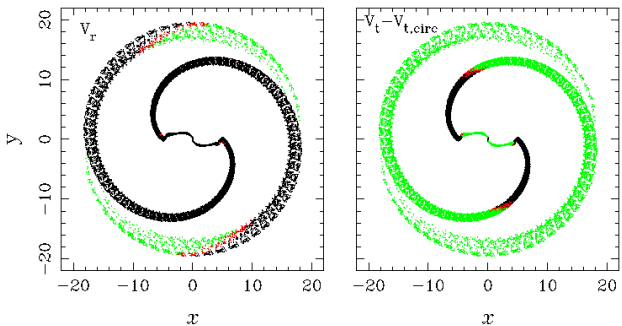} 
\caption{As in Fig. \ref{fig:rR-vrvt-xy} but the velocities are now
  calculated along the spiral arms  
  of a D model with $\epsilon$ = 0.3 and $r_l$ = 5. 
}
\label{fig:sp-vrvt-xy}
\end{figure}

\begin{figure} 
\centering
\includegraphics[scale=0.44,angle=-90.0]{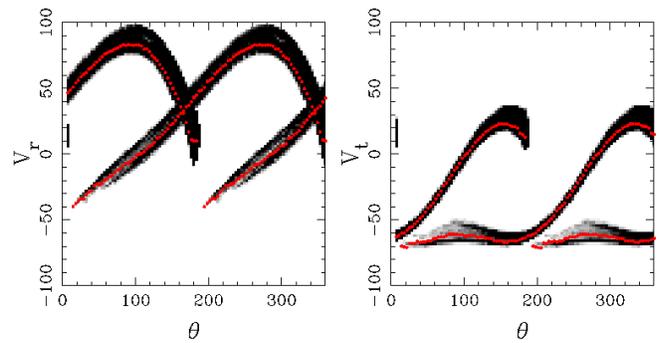} 
\caption{Radial (left panel) and tangential velocities (after
  subtraction of the circular velocity -- right panel)
  along the spiral of a D model with $\epsilon$ = 0.3 and $r_L$ = 5. 
  The red curve gives the median values.
}
\label{fig:sp-vrvt-theta}
\end{figure}

\begin{figure*} 
\centering
\includegraphics[scale=0.65,angle=-90.0]{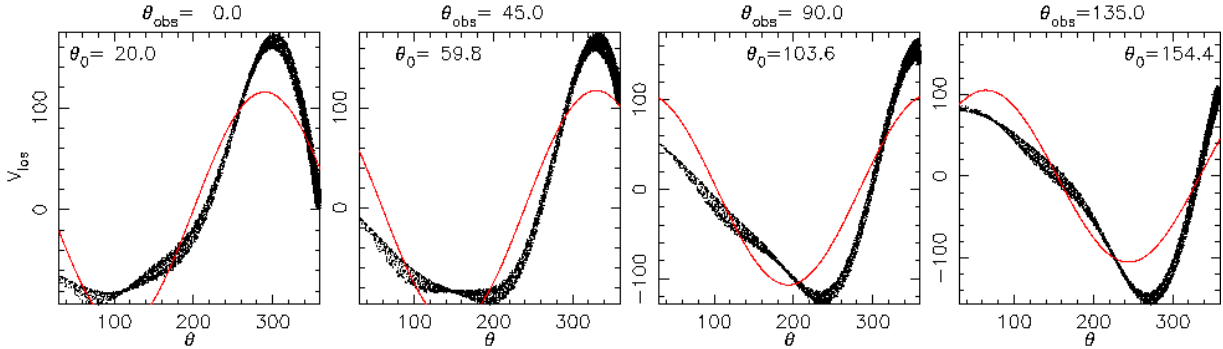} 
\caption{Line-of-sight velocity for points along one of the spiral
  arms for the model shown in Figs.~\ref{fig:sp-vrvt-xy} and
  \ref{fig:sp-vrvt-theta}. Each panel corresponds
  to a given viewing angle, whose value is given in the top of the
  panel. All angles are in degrees. The red lines are the best fitting
  sinusoidal curves, which in all these cases are poor fits. 
}
\label{fig:sinfits}
\end{figure*}

Let us now consider a case with a much stronger relative
non-axisymmetric forcing, which, in agreement with what was found in
Paper III, has a spiral morphology. We will use a D model 
with  parameters $\epsilon$ = 0.3 and $r_L$ = 5, and we plot the
radial and tangential velocities in Figs.~\ref{fig:sp-vrvt-xy} and
\ref{fig:sp-vrvt-theta} (left and right panels,
respectively). Material on orbits emanating from the vicinity of $L_1$
or $L_2$ moves outwards with increasing radial velocity for the first
roughly 80$^\circ$, then the radial velocity reaches a maximum and
starts decreasing. The material still moves outwards, however, up to a
winding of 
roughly 260$^\circ$ at which point the radial velocity changes sign
and the motion changes to inwards. Similar information for the tangential
component of the velocity is given in the right panels of these
figures. Material will move from $L_1$ or $L_2$ with a tangential velocity which
is above the circular velocity until it has described roughly
80$^\circ$ and then will continue with a smaller tangential
velocity. After about one rotation, however, the tangential velocity
will stay at a roughly constant value.

From the cases described above, and from a number of others which we
have run but do not show here, we see that there is a continuity
between the kinematics of material along $R_1$ rings and kinematics
along spirals, with, nevertheless, clear quantitative differences. 
For example, for stronger relative non-axisymmetric
forcings the radial velocity changes sign much
further along the arm i.e. further away from the Lagrangian point from
which the arm emanates. Let us call the angle at which this change
occurs $\theta_{cs}$. It is measured from the Lagrangian point from
which the manifold emanates in the sense in which the particles are
moving. The A model 
described in Figs.~\ref{fig:rR-vrvt-xy} and \ref{fig:rR1-vrvt-theta} 
has $Q_{t,L_1}$ = 0.038 and $\theta_{cs}$ = 80$^{\circ}$, while the A model
described in Fig.~\ref{fig:rR1-vrvt-theta-strong} has $Q_{t,L_1}$ = 0.072 and
$\theta_{cs}$ = 90$^{\circ}$. A D model with
$\epsilon$ = 0.25 and $r_L$ = 5. has $Q_{t,L_1}$ = 0.515 and
$\theta_{cs}$ = 230$^{\circ}$, while the D model with $\epsilon$ = 0.3 and
$r_L$ =
5 (Figs.~\ref{fig:sp-vrvt-xy} and \ref{fig:sp-vrvt-theta}) has
$Q_{t,L_1}$ = 0.618 and $\theta_{cs}$ = 255$^{\circ}$. 
From these and other models we see that there is 
definitely a trend, but it is beyond the scope of this paper to see
whether this is a tight correlation and whether this is model dependent.
Similar trends can be found also for the tangential velocity. To what
extent this result can be modified by self-consistent calculations is
still unclear.

Another difference can be seen by comparing
Figs.~\ref{fig:sinfitrR} and \ref{fig:sinfits}.
The former shows that for both inner and outer rings and for all
values of the viewing angle the line-of-sight velocity can be well
fitted by a sinusoidal curve. Fig.~\ref{fig:sinfits} gives the  
line-of-sight velocity for the spiral case. The velocities are now
measured along 
the spiral arm. One sees clearly that in this case a sinusoidal is a
bad fit to the data and should not be used to find the galaxy major
axis. 

\section{Pattern speed prediction}
\label{sec:omegap}

The pattern speed, i.e. the angular velocity of the bar, conditions
strongly the dynamics within a barred galaxy. Unfortunately it can not
be directly observed and it is thus necessary to resort to indirect,
and often rather convoluted methods, to obtain its value.
Orbital structure studies of
barred galaxies have shown that a bar can not extend beyond its
corotation \citep{Contopoulos80}. A similar result was reached from
the calculation of the self-consistent response to a bar forcing
\citep{Atha80}. These two results are very useful, since they set a
lower limit to the corotation radius, which can thus not be smaller than
the bar length, but give no information on an upper limit. A two-sided
constraint, i.e. one limiting the Lagrangian radius from both smaller
and larger radii,
has been found from the comparison of shock loci in hydrodynamic
simulations and dust lane loci observed in barred galaxies. The
possible range for the Lagrangian radius is thus found to be
$r_L = (1.2 \pm 0.2)a$,
where $a$ is the bar semimajor axis \citep{Atha92b}. This has proved to
be the tightest constraint to date and compares well with 
observational results \citep{Elmegreen96}, the most extensive of
which are from the application of the Tremaine-Weinberg method
\citep{Tremaine.Weinberg84} to bright early type galaxies
(\citealt{AguerriDC03, GerssenKM03, Corsini.Debattista09} and
references therein). Yet the 
observational errors for all methods are rather large and 
more precise determinations would be desirable \citep{Corsini.Debattista09}. 
We will examine here whether, and in what cases, our work could
provide some input on pattern speed values. Indeed, from all the above and 
from Papers I, II and III  it is clear that the position of the
Lagrangian points is intimately linked with the morphology and thus
can perhaps be estimated from it. 

If the galaxy has a bar and no spiral, then the potential is symmetric
with respect to both the bar major and minor axes and the $L_1$ and
$L_2$ are on the direction of the bar major axis. If its morphology is
$rR_1$ and the inner and
outer rings join, then the $L_1$ and $L_2$ are located at the
position where the two rings join. If they do not join, but have a
considerable distance between them, then the $L_1$ and $L_2$ can be 
stable, as discussed in Section 5 of Paper III. 
In this case, the Lagrangian points are situated in between
the extremity of the inner ring major axis and the dimple of the outer
ring. Taking it half way between the two is a good approximation.
If the galaxy morphology is $R_1R_2$, the $L_1$ and $L_2$ can be
localised in the same way as for the  $rR_1$. 

In cases where the galaxy has an $R_1$ outer ring with a clearly
visible dimple but no inner ring, or the inner ring is not clearly
delineated, the Lagrangian points should be in between the
dimple and the tip of the bar major axis, but their exact location is
difficult to pinpoint, unless a full model of the galaxy is made. The
most difficult case is if the inner ring does not exist, or is very
badly delineated, and at the same time the outer ring shows no clear
dimples. In such a case the Lagrangian points should be on the
direction of the bar major axis, between the end of the bar and the
outer ring. Locating them half way between the two is the best guess,
but has a large relative error. Even in such cases, however, this
method can be useful since it gives some, albeit rough, 
estimate of the pattern speed very fast, simply from a galaxy image
and without any need for kinematic observations.  

In cases of barred galaxies with no rings and a relatively weak spiral
structure emanating from the end of the bar, then the $L_1$ (or $L_2$)
is located at the tip of the bar, where the arm joins the bar.  
In some cases this may be somewhat difficult
to determine precisely in observations, since the distinction between
the end of the bar and the beginning of the arm is not always easy to
make. Nevertheless the error 
this entails is not very large. Cases where the spirals do not come
directly out of the tip of the bar major axis are less clear cut. 

$R_2$ morphologies with a spiral between the inner
ring (or bar) and the outer ring, as ESO 325-28 (see third column of
Fig.~2 in Paper IV), can be considered in the same way as simple
spirals. Cases with only an $R_2$ ring do not present sufficient
morphological characteristics to allow measurements of the $L_1$ and
$L_2$ accurately. And of course nothing can be said from morphology
alone about cases with no spirals and no rings.

An interesting peculiarity can happen in the case where the
contribution of the spiral to the forcing is considerable. Contrary to
the bar-only case, in bar-plus-spiral cases
the iso-effective potential curves will not be symmetric with respect
to the bar major and minor axes, since the spiral contribution 
lacks that symmetry. The Lagrangian points $L_1$ and $L_2$ would
then not be located on the direction of the bar major axis, but at an
angle to it. Stronger spirals will result in larger angles and
trailing spirals will make clockwise displacements (leading spirals
anticlockwise). This means that, if the spiral is sufficiently strong
compared to the bar, corotation will not be exactly at the position 
where the bar joins the arm, but can be somewhat further down the
arm. In such a case and at that location, the position of the dust
lane will shift from the inside to the outside of the arm. Such a shift has
indeed been observed in the SE arm of NGC 1365
\citep{Ondrechen.VanDerHulst89,JorsaterVanMoorsel95}, thus
bringing corotation to $1.4a$, where $a$ is the bar semimajor
axis. This value is in good agreement with the value found from the
hydrodynamical modelling of NGC 1365 \citep{Lindblad.LA96, ZanmarSSWW08}, and is
within the bracket set by the gas flow simulations of \cite{Atha92b}, namely 
$r_L = (1.2 \pm 0.2)a$.

Once the Lagrangian points $L_1$ and $L_2$ have been clearly located,
it is in principle 
straightforward to obtain the pattern speed. From the rotation curve
one can obtain the curve $\Omega = \Omega (r)$ and its numerical value
at $r = r_{L}$ is the pattern speed $\Omega_p$. In cases with strong
bars, however, the position-velocity curves along and perpendicular to the bar
major axis differ considerably (see e.g. \cite{DuvalAthanassoula83}
for NGC 5383). Yet, and as can be seen e.g. from Fig. 16 in the afore
mentioned paper, this difference is substantial mainly in the inner
parts and considerably less so at the corotation region. Furthermore, if
a sufficiently accurate velocity field is available, a yet more
precise value of the pattern speed can be calculated from the rotation
curve, since this is calculated from an azimuthal average of the
velocities.  

\section{Further discussion}
\label{sec:discussion}

\subsection{Self-consistency and spiral structure formation}
\label{subsec:selfspiral}

As discussed in Paper I (see particularly figure 6 there) the arm or
outer ring forms gradually, from 
the Lagrangian point outwards. Any mass element needs some time to
leave the vicinity of the $L_1$ or $L_2$ and to complete half a
revolution around the galactic centre, and this time depends on 
the model. For strong bars this time is short, while for weak ones it
is considerably longer. In Paper III we gave examples where 
this time varies roughly between 2 and 10 bar rotations. Furthermore,
as shown in figure 6 of Paper I, the speed of formation depends
strongly on the position along the arm. The part of the arm nearest to
the Lagrangian point takes the longest to form and this time
decreases with increasing distance from the arm origin.  

In Paper III we showed that the bar strength plays a crucial role in
determining the morphology and we gave a value which is a rough
threshold between rings and spirals. Thus, bars less strong than this
threshold will result in $rR_1$ or $rR_1'$ morphologies, and stronger
ones in spirals. However, the time dependence discussed in the
previous paragraph argues that, in 
a fully self-consistent model in which time evolution is taken into
account, the limits between the different morphologies will be 
shifted with respect to what we found in Paper III for simple rigid
bar potentials where self-consistency is neglected. 

To better understand this, let us consider a bar which has a
strength somewhat below the limiting value necessary for the manifolds
to have a spiral morphology, i.e. a bar strength given in Paper III as
corresponding to an $R_1$ or an $R_1'$ outer ring. As we showed in
Paper I (see particularly figure 6 there and corresponding discussion), the
manifolds grow gradually from the $L_1$ (or $L_2$) outwards. At the
initial stages of ring formation, only a segment of the ring has 
formed, so that the ring is not closed and looks
like a short spiral arm. Its contribution, however, to the forcing
is not taken into account in non-self-consistent simulations. 
So the evolution will give rise to the $R_1$ or
$R_1'$ morphology which corresponds to the bar forcing. 

On the other hand, in self-consistent simulations this segment will
contribute an extra forcing which will foster a manifold with a spiral
shape. The joint forcing of this ring segment and of
the bar will drive manifolds which have a different shape than those
that a bar only would drive, and are, in particular, more spiral-like.
The resulting new segment, together with the bar, will give
another manifold shape, etc, so that the final morphology can, in
this case, be a trailing two-armed spiral. Thus, if the bar is
sufficiently near the dividing line, it may have manifolds of a ring
shape in non-self-consistent simulations and of spiral shape in
self-consistent ones. As a result, the limiting bar strengths which divide the
different types of manifold morphologies will be pushed downwards, to
lower values of the bar strength. Fully self consistent simulations
would be necessary in order to find out how much this shift will be.

\subsection{Nature of the relevant orbits and differences between
  rings and spirals}
\label{subsec:chaos}

As already discussed in the previous papers of this series, the orbits that
constitute the building blocks of spirals and rings are chaotic. Yet they
are spatially well confined by the manifolds, which can be thought of
as tubes guiding the orbits. They can thus form features as narrow and
as well defined as the spiral arms and rings observed in barred
galaxies. For this reason, we propose to name this type of chaos  
`confined chaos'. 

Intuitively, using chaotic orbits may seem a rather
poor way of building narrow structures, since, given a sufficiently
long time, chaotic orbits tend to cover all the phase space 
that is available to them \citep[e.g.][]{Contopoulos02}. Yet the time
required for this may be quite long, sometimes much longer than the time
scale of the problem at hand, or even the Hubble time. Moreover, the
phase space available 
to chaotic orbits is limited by the regions of phase space occupied by
regular orbits and thus in some cases its projection on configuration space
may be spatially very limited. Thus confined chaotic orbits may well  
contribute to the formation of galactic structures, even to ones which
are narrow features in configuration space. Examples of
such a case are the chaotic orbits guided by the manifolds in the
outer parts of a strong bar. Although chaotic, these orbits are confined by
the inner manifold branches and can thus outline the shape of the bar
and its outer isodensities \citep[see also][]{PatsisAQ97, VoglisHC07}

It is thus necessary to keep a broad view of what constitutes a
building block of a galaxy. Indeed, 
we need to consider in all cases not only the
regular orbits trapped around the stable periodic orbits, but also
confined chaotic orbits. The latter have received little attention
so far and could bring new light to many galactic dynamics problems.

Let us now discuss separately the orbits in rings and in spirals.
In the case of rings, a particle caught by a manifold stays there 
indefinitely\footnote{This is true in theory and for rigid
potentials. For real galaxies, see Sect.~\ref{subsec:time}}. This may seem in
contradiction with the fact that the orbit is chaotic, but in fact it
is not. Chaotic orbits indeed occupy all available phase space, but
the projection of this space on to configuration space 
can be severely confined by regions with regular
orbits. For example, if we take a chaotic orbit following the path
outlined in panel c of Fig.~\ref{fig:circring}, the configuration
space available to it is confined from the inner part by the 
stable periodic orbits around $L_4$, by the regular orbits
trapped around them and by the curve of zero velocity of the same
energy as the orbit. Similarly, from outside the confinement comes
from the stable periodic orbits at larger radii from the centre and
from the regular orbits trapped around 
them. So the chaotic orbits trapped by the manifolds do occupy all the 
available phase space, but in this case the projection of the
available phase space onto configuration space is of very limited
extent. The important point to 
underline is that, in this case, the form of the available
configuration space is just appropriate to reproduce the
observed shapes of $r$, $R_1$, $R_2$, or $R_1R_2$ rings.

The case of spirals can be somewhat different. Let us again consider
the case we have worked with so far, namely 
a rigid rotating bar potential. Again a particle which is trapped in 
a manifold will stay there indefinitely (in theory at least). But now the
manifolds `widen up' as they move away from the region around the
Lagrangian point from which they emanate. This means that after one or
more revolutions around the galactic centre (in the rotating frame of
reference) the manifolds may be too wide to outline spiral arms. This can
in practise limit the radial extent of the arm, but this limit is not
sharp -- contrary to the inner and outer Lindblad resonances which are
sharp limits for non-driven stellar density waves
\citep[e.g.][for a review]{Lin67}. The orbits will then, while still
being in the manifold, contribute to the axisymmetric background
rather than to the spiral arm. This has several implications. 

The first one concerns response simulations in purely stellar
cases with a rigid potential. \cite{Schwarz81} performed such
simulations in which the bar was initially grown gradually and then
kept at a constant amplitude and he found that the response was a spiral
during the time when the bar was growing, but no spiral could be seen after
that. This is indeed what our theory predicts and there is thus good
agreement between our theory and these simulations. A second implication is
that a particle has a `useful life', during which
it will contribute to the arm and after which it will contribute to the
disc density. How this affects the life-time of spirals and whether
these are long- or short-lived is discussed in Sect.~\ref{subsec:time}.

\subsection{Time evolution. Are spirals and rings long-lived, or short-lived?}
\label{subsec:time}

In a galaxy we do not observe the manifolds themselves, but the
material that is confined by them. Indeed, the manifolds 
are only the building blocks and thus may exist, but trap few or no
orbits, in which case the corresponding structure will be very faint
or even not 
visible in the galaxy. This is similar to the periodic orbits, which
are the building blocks of bars and which need to trap regular orbits
around them for the bar to form. In other words, the existence of a
manifold is a necessary, but not a sufficient condition for the
corresponding galactic structure to form. Thus, manifold theory
without any information on the formation of the galaxy can
answer questions concerning the shape of a spiral or a ring, but
can give little or no information on its amplitude. 

In a purely stationary model, material can not move in or out of a
manifold. In a galaxy, however, a gas cloud or a star can do so
after a collision (e.g. if it encounters another gas
cloud), or if the bar potential or pattern speed change. Most of the
material is expected to be trapped in a manifold as the bar
forms and grows. Thus, in principle at least, information on
how much material is contained in the manifold could be obtained by
calculations including time evolution, so that the trapping of the
material in the manifolds can be followed (Paper IV).

Are spirals in our theory short-lived or long-lived? Spirals will
exist as long as there is sufficient material guided by the
manifolds. As already discussed, this
material comes from the bar region, is fed to the vicinity
of the Lagrangian points by the inner stable branches of the manifolds
and from there moves outwards along the arms to 
%. The manifolds broaden considerably as we move outwards, so
%that this material will, after some time, 
get dispersed in the outer parts of the galactic disc, due to the
broadening of the manifolds. Thus the 
spiral will survive only as long as the reservoir (i.e. the outer
parts of the bar and the vicinity of the Lagrangian points) is capable
of providing fresh material. When this  dries up, the spirals will
fade out although the manifolds still exist. In this sense, the
spirals can be considered as relatively short lived. 

However, it is now well established from $N$-body simulations that
bars evolve with time, getting longer and stronger while slowing
down \citep[][etc.]{Debattista.Sellwood.00, Atha.Misiriotis02, Atha03, Atha05,
Valenzuela.Klypin03, ONeill.Dubinski.03, DebattistaMCMW06,
MartinezVSH06, BerentzenSMVH07}. The Lagrangian points move outwards
and at the same time 
the outer parts of the bar occupy a different region of the galactic
disc. Thus the mass reservoirs get replenished and they can feed mass
to the spiral for longer times, although the form of these spirals
will evolve with time (since the bar strength and $r_L$ change). 
Further work (Romero-G\'omez et al, in preparation) is necessary in
order to estimate the bar evolution that can maintain a long-lived
spiral, and whether any fine-tuning is necessary for this. So at this
stage we can only present this as a 
possibility, but lack solid proof. To summarise, our theory predicts
short lived spirals if the bar does not evolve, or evolves little, but
could in principle predict relatively long lived but evolving spirals
if the bar grows 
sufficiently with time. In no case do we get a long-lived stationary
spiral. If for some reason the bar growth stops, the spiral will not
survive for long and will not be visible after the reservoir has dried
out. The rate at which new 
material needs to be provided depends on the instability of the
Lagrangian points, i.e. on the strength of the bar. 

The situation is different for rings. Fig.~\ref{fig:circring}
shows the motion in an $rR_1$ morphology. In such cases, once material
is trapped by the manifolds, it can stay there indefinitely, if the
potential does not change with time and if other forces (such as
collisions, or dynamical friction) do not force it
to leave the arm. Furthermore, if the potential evolves very slowly,
the rings could follow it by adiabatically changing their shape. Thus,
rings should be more long-lived than spiral arms and do not
rely on strong secular evolution for their existence, but may adapt their
properties to the bar evolution. 

\subsection{Comparison with other spiral structure theories and with
  observations} 
\label{subsec:othertheories}

\begin{figure*}
\begin{center}
\includegraphics[scale=0.33]{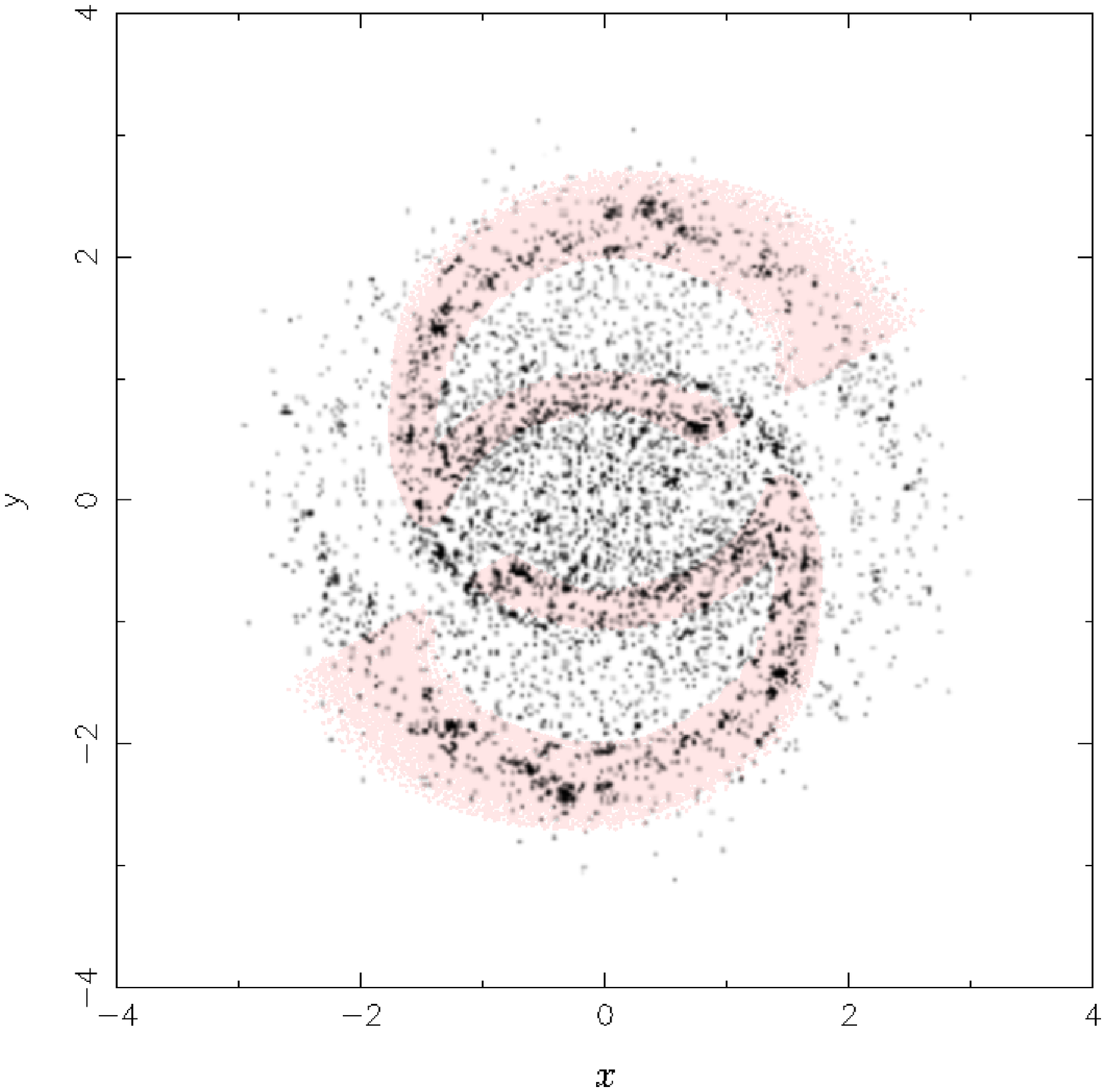}
\includegraphics[scale=0.33]{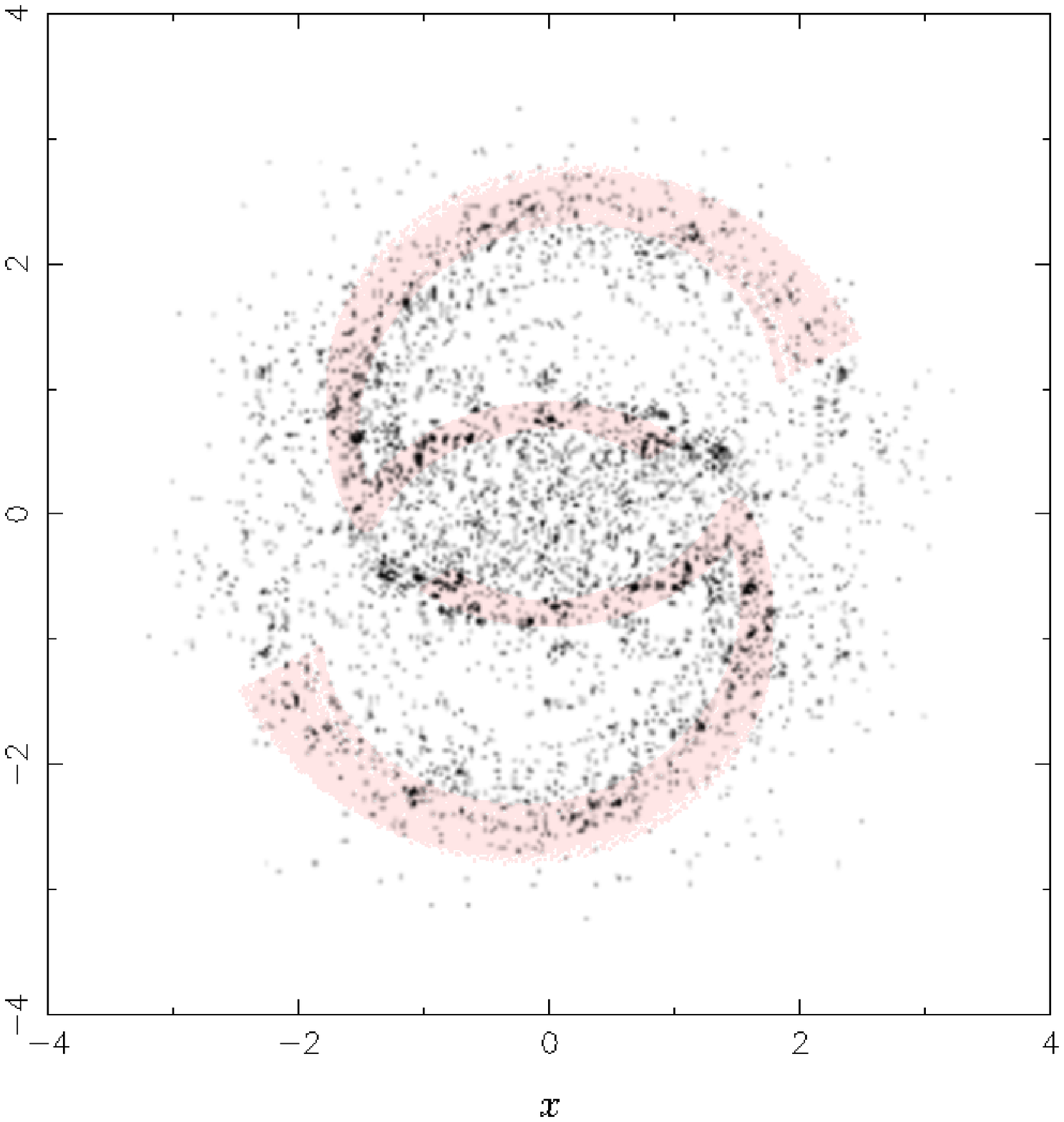}
\includegraphics[scale=0.33]{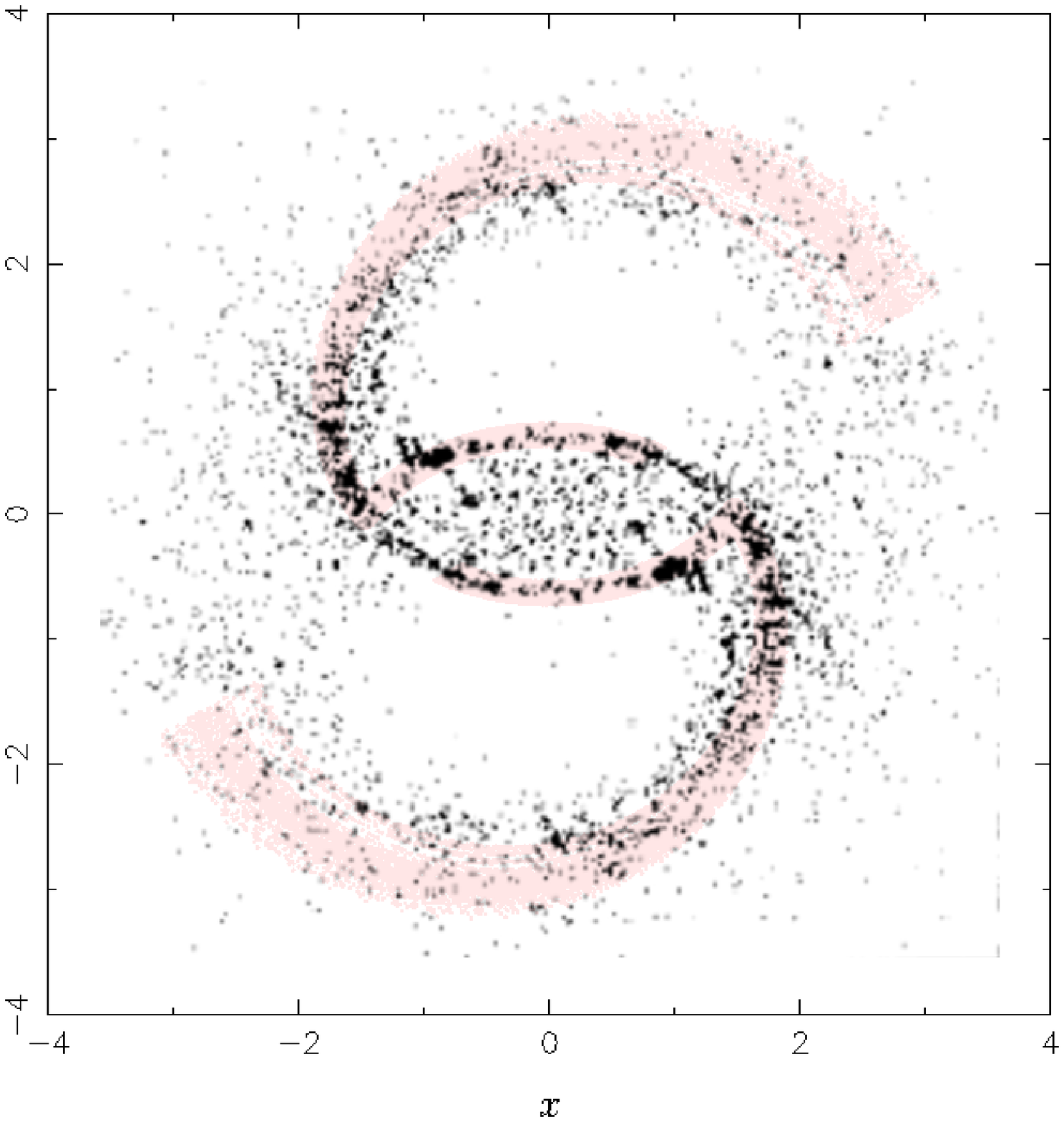}
\end{center}
\caption{Comparison of the loci of gaseous manifolds to the gas
  response. The manifold loci are given in pink and the positions of
  the gas clouds in the response simulations in black. The latter are
  taken from \citet{Schwarz84}. We make a comparison for three models
  with bars of different strengths, the
  bar strength increasing from the left to the right in the
  figure. 
} 
\label{fig:compschwarz}
\end{figure*}

According to our theory the building blocks of spirals and of rings in
barred galaxies are chaotic orbits guided by the manifolds associated
to the $L_1$ and $L_2$ Lagrangian points. This, however, is not the
only alternative, and most spiral structure theories are based on
the density wave concept initially introduced by
\cite{Lindblad63}. Grand design spirals can thus be driven e.g. by
companions, or by bars, or perhaps be modes relying on some amplification
mechanism. Flocculent spirals can be due to local swing
amplification, while self-propagating star formation has also been
proposed. Several origins for spirals are thus possible depending on the case  
(existence and relative orbit of the companion, existence of a strong
bar, etc), so that spirals in different galaxies may have different
origins. Furthermore, even in a single galaxy more than one mechanism can be at
play. The properties of the resulting spiral would then result from
the self-consistent interaction of all the spirals of different
origins. Thus for example, the density from orbits trapped by the
manifolds can help 
drive a response which will include orbits of energies other than the
ones concerned by the manifolds discussed here, so that the resulting
spiral will be somewhat different from what would be expected 
from the manifold theory alone.

Similarly, different alternatives can
occur when we consider the fate of a gas cloud falling on a manifold
arm. If the potential is evolving, i.e. the bar is growing, or if the
motion of the cloud is perturbed by a collision with another cloud, it
can be trapped by the manifold and continue its motion along the
arm. Whether this will trigger star formation or not can not be
answered here. On the other hand, if the cloud is not trapped by the
manifold, it will cross the arm and its fate
will be as described by \citet{Roberts69}, i.e. there can be star
formation and an age gradient roughly perpendicular to the arm. 

Our theory leads naturally to a preference of two-armed grand-design
trailing spirals in barred galaxies (Paper IV), in good agreement with
observations. Other theories and other 
circumstances, however, may result in different spiral properties
\citep[see][for reviews]{Toomre77, Atha84} and not all theories can
show a preference for trailing two-armed spirals. Resonances aside, the
WKBJ density wave dispersion relation does not discriminate between leading
and trailing spirals, both being equally possible, while the WASER
feedback cycle involves three waves, long or short but always
trailing. Driven density waves have a given sense of winding depending
on the driving agent. Namely, a growing bar will drive a
trailing two-armed spiral, a directly rotating companion will drive
mainly two-armed trailing spirals and a retrograde satellite a one-armed
leading spiral. The swing amplifier feedback cycle relies on three
waves, the two within corotation being one leading and the other
trailing \citep{Toomre81}. The ratio of their amplitudes depends on
the amplification factor. Within corotation, the superposition of
the two waves will give an interference pattern (see e.g. figure 12 in
\cite{Toomre81}, or figure 31 in \cite{Atha84}) and Fourier analysis
should reveal the existence of the two 
components, provided of course the amplification factor is not too
large. A similar statement can be made for our theory. If the ratio
of the amount of material trapped in the unstable to the material trapped
in the stable manifold is not too large, interference
pattens will be clearly visible and the leading component will be
detectable by Fourier analysis. In fact, this has been observed in 
a number of nearby galaxies \citep{Atha92c, PuerariBEFE00}. On the
other hand if the ratio is very large, then only the 
trailing component can be detected. Unfortunately these properties
alone will not allow us to distinguish between the swing amplifier and
the manifold origin of a spiral, since both theories can account for
the presence or absence of a leading component and of interference patterns.

The bar strength affects the properties of the spirals in a different
way in driven density waves and in our theory. In the former, and
unless yet unknown nonlinear effects can modify this, the strength of
the bar affects the amplitude of the spiral in the sense that stronger
bars will drive higher amplitude spirals. In fact in a linear theory,
the spiral amplitude is proportional to the bar strength. On the other
hand in our 
theory the bar strength affects the pitch angle of the spiral, but not
necessarily its amplitude, since the strength of the bar will change
the manifold shape but may not affect its amplitude which depends only
on how much mass the manifold traps. As
discussed in Paper IV, it is not straightforward to test this fully
with observations. Existing results, however, give us some crucial
information. Thus, \cite{Seigar.James98} and \cite{SeigarCJ03} find that
there is no correlation between bar and spiral strength. The
question was revisited by \cite{ButaKESLEPB09}, using a sample of 177
galaxies and a different method for measuring the bar and the
spiral strength. They find a very weak trend, with a linear correlation
coefficient of 0.35 and a Kendall tau rank coefficient of 0.22. These
numbers are so low that the existence of a trend is
questionable. The lack of any trend or correlation was
confirmed by \cite{DurbalaBSVM09}, using Sb - Sc galaxies from the
AMIGA sample of isolated galaxies
\citep{VerdesMSLLEGSV05}. \cite{SaloLBK09}, however, argues again in
favour of a correlation. The observational situation is thus still
unclear. The existence of a correlation
would be in good agreement with the linear density wave theory, while
the lack of such a correlation would argue against it, at least in
its current linear form. On the other hand, the theory we propose
here, can not give definite information as to whether such a
correlation should exist or not, unless further work concerning the
trapping is made. Thus, the existence or the absence of such a
correlation can neither confirm nor reject it.

Compared to density wave theory, our theory has a number of further
advantages. It can explain the formation not only of spirals, but also
of rings, and reproduce their main morphological properties. The
continuity between rings and spirals follows naturally from
our work, while it can not be explained by density wave theory. Our
theory can also reproduce the shape of spiral arms 
such as those of NGC 1365, NGC 1300, or NGC 1566 \citep[see figures in
  e.g.][]{VeraVDPC01}, which `fall-back' towards the
bar region or towards the opposite arm (Sect. 2 of Paper IV), while
this can not be reproduced by density wave theory.  

This difference between the arm shapes as predicted by our theory and
by density wave theory has a corollary. Namely, density wave theory
predicts that velocities will be outwards in all the spiral arm region of
a barred galaxy, since this region is beyond corotation. On the other
hand, for our theory, these velocities will be outwards for most of
the spiral extent, but will turn to inwards in the outermost part of
the spiral, where the arm starts falling back towards the bar, or
towards the other arm (see Sect.~\ref{sec:Kinematics}). These parts,
however, are very low density, particularly for strong bars, so
kinematical observations will not be easy to make. Furthermore, it is
unclear at this stage how much this result will be affected by a self
consistent estimate of the potential, although this may not be so
important for such low density parts.   

A further difference between the density wave theory and our theory,
at least in its present non-selfconsistent version, could be found in the
effect of the velocity dispersion. According to density wave theory,
this will influence strongly the pitch angle of the arms, as has been
shown in the local theory, in global modes and in response spirals
\citep{Kalnajs70, Lin.Shu71, Atha78, Atha80, Bertin.LLT89,
  Hozumi03}. As a stellar 
population becomes older, it will emit stronger in the IR and the NIR
and less strongly in e.g. the blue wavelengths. At the same time
its velocity dispersion will increase
\citep[e.g.][]{Binney.Tremaine08}, so that, all other parameters and
properties remaining constant, its spirals should become considerably
more open. Thus, according to the density wave  
theory, there should be a considerable difference between the winding
of spirals as observed in the blue and in the NIR (for quantitative
estimates see the above mentioned references). On the contrary, in
our theory stars of different ages will trace the same spiral arm,
i.e. they will be guided by the same manifold, if they have the same
energy. Thus the only way for spirals viewed in different wavelengths 
to have a considerably different winding, is for the distribution of
the energies of the stars, and particularly the mean energy, to depend
strongly on 
age. Although a quantitative estimate of this dependence is not
available at present, we wish to underline this as a useful subject for
future work, since it could shed some light on the differences between 
spirals due to a density wave and spirals due to our theory. 

It should also be kept in mind that the corresponding observational
test is not easy, since it is difficult to disentangle the effect of
the velocity dispersion from that of other parameters. Yet the
images of the S4G sample \citep{Kartik10} show that there is
little, if any, difference between the winding of the arms in the blue
and at 3.6$\mu$m \citep[][and private communication]{Buta10},
certainly much less than predicted by e.g. the estimates in \cite{Hozumi03}. 
Furthermore, \cite{Buta10} found that galaxies at 3.6$\mu$m tend 
on average to appear ``earlier'' in type than in the blue, by approximately
one stage. This argues against a statistical `opening' of the arms as
one goes from blue to mid-IR. Although further work is necessary in
order to substantiate the above statements, these results show that there
might be a potential problem with density waves in this respect, which
the manifold theory would not suffer from.  

In our theory, the bar and the spiral have the same pattern speed, 
contrary to some other theories, where more than one pattern
speed can be present in the same disc galaxy \citep{TaggerSAP87,
  Sellwood.Sparke88, SygnetTAP88, Masset.Tagger97}, so that the spiral
can rotate with a different pattern speed from that of the bar. Unfortunately
it is very difficult, if at all possible, to distinguish between these
cases observationally. One possibility would be to check the position
of the dust lane along the arm with respect to the arm centre,
i.e. to check whether it is on the trailing or the leading side of the
arm. Similarly one could check where the radial motion switches from
inwards to outwards. Both, however, are very difficult to
pinpoint. The task is further complicated by the fact that if the
spiral potential is a sizeable part of the total, then the Lagrangian
points are not on the bar major axis, but they move to a position
somewhat further out along the arm. Thus, e.g. in NGC1365, the
crossing of the dust lane from one side of the arm to the other does
not necessarily imply that the spiral has a different pattern speed
from that of the bar. Indeed, as discussed in Sect.~\ref{sec:omegap},
our theory, in which the bar and the spiral have the same pattern
speed, can also explain this shift. 

Work along lines similar to ours has also been carried out by others.
\cite{Danby65} argued that bar orbits departing from the
vicinity of the unstable Lagrangian points play an important role in
the formation of the spiral arms. \citet{KaufmannContopoulos96} 
linked chaotic 
orbits to the presence of spiral arms. This was further developed by 
\citet{Patsis06} with the help of response calculations in the
potential of NGC 4314 (as calculated by \citet{QuillenPG94}) and by
\cite{VoglisSK06} who used a potential from an $N$-body simulation.
Manifolds were specifically referred to, although in a quite
different way from that used in our work, by the late Prof. Voglis and his
collaborators \citep{VoglisTE06}. These authors consider only spirals
--and not rings -- and, as underlined by \cite{TsoutsisKEC09}, they do
not consider the swarm of orbits trapped by the manifolds, but the
loci of their manifold apocentra. Due to the 
differences between our work and that of Voglis and his collaborators,
our theoretical results and our comparisons 
with observations can not be straightforwardly extended to the Voglis
et al. studies.  

\subsection{Comparison with response simulations}
\label{subsec:othersimu}

We can also compare our results with those of simulations. Since our
calculations are not self-consistent, we will make comparisons only
with response simulations. Such comparisons can be even quantitative
for simulations such as those of \cite{Schwarz79, Schwarz81,
  Schwarz84}, where the gas clouds are modelled as colliding
particles. We employed a similar technique when 
calculating the orbits in our manifolds (see Paper III for a technical
description of the method) and we make a comparison with Schwarz's
results in Fig.~\ref {fig:compschwarz}. The three panels correspond to
three models of different bar strength given in the lower panels of
figure 1 of \cite{Schwarz84}. The positions of the gas clouds in the
simulation are given in black and the manifold loci in pink. 

We see that the agreement is very good, particularly if we take into
account that our treatment of collisions is very similar, but not identical
to that of Schwarz, because we do not have a population of colliding
clouds. Also, and for the same reason, the number of collisions we use
is somewhat arbitrarily chosen. In the examples in Fig.~\ref {fig:compschwarz}
we took one collision per half bar rotation, but roughly equally
good results can be found if the collision rate is doubled.

This argues that the orbits in Schwarz's simulations are guided by
manifolds such as those we discussed in our series of papers. 
This argument is further strengthened by further good agreement found
between our results and those of Schwarz. For example,
\cite{Schwarz81, Schwarz84} finds that in his simulations R1 rings do
not form in cases with strong bar forcings. Similarly, in his
simulations he finds that stronger bars drive more open spirals than
weaker bars. Both these simulation results have now been understood and
explained with the help of our work (Paper III) and are a natural
consequence of our theory.

In Papers II and III we showed that in the $R_1R_2$ morphologies,
material from the outer part of the bar fills in first the parts of
the manifolds that correspond to the $R_1$ morphology and only
afterwards the $R_2$ part. In other words, the $R_1$ part of the
morphology forms first and is followed by the $R_2$. This was indeed
first seen in the response 
simulations of \cite{ByrdRSBC94}, who ran a large number of
simulations very similar to those of Schwarz aiming to understand the
evolution and morphology of the gaseous distributions in barred
galaxies. This fact is also in good agreement with
observations. Indeed, the Catalogue of Southern Ringed Galaxies
\citep{Buta95} has revealed the existence $R_1R_2$ morphologies with $R_1$
rings which are weak in blue light and very strong in the $I$ band. A good
example is IC1438.  

More comparisons are of course necessary in order to fully establish
the fact that the motions of particles in response simulations are
guided by manifolds as the ones described in this series of papers.
Nevertheless, the morphological agreement in Fig.~\ref {fig:compschwarz} and in
other models (not described here), as well as the other agreements
presented in this series of papers, make this
very plausible. If the correspondence between manifolds and simulation
results is indeed established, the good agreement between response
simulations and observations found for many galaxies
\citep[e.g.][]{ButaCB99, Salo.RBPCCL99, Rautiainen.SL05,
  TreuthardtSB09} will also be
found when comparing the observations to the manifold loci in the
appropriate potentials. 

\subsection{Future work}
\label{subsec:future}

In this series of papers we developed some basic aspects of
our theory and compared it to observations. We used simple rigid
potentials in which we calculated manifolds and orbits. This allowed
us to get a good understanding of the underlying dynamics and its
applications and showed that the theory we propose can play a crucial
role in explaining the formation and properties of spirals and rings
in barred galaxies. We will outline here a number of avenues for
future research which should contribute further to assessing
our theory and to extending it to other types of spirals and other
configurations.

We have so far developed the theory only for the case of saddle point
Lagrangian points. Yet the contribution of confined chaos to the
morphology and structure of galaxies extends much further than this.
For example, in Sect.~4.2 of Paper IV we saw that the $L_4$
and $L_5$ Lagrangian points can become complex unstable for very strong
Ferrers bars. It is thus necessary to understand whether and in which
way the corresponding manifolds can, in such cases, contribute to
galactic morphology. Also, most families of periodic orbits include at least
some energy ranges with unstable orbits, whose orbits can not trap material.
In such cases as well, confined chaos and manifolds may 
contribute to create 
morphological features. Considerable theoretical work is still
necessary to fully explore all these theoretical aspects.
  
So far we have used three types of bar potentials, the Ferrers
\citep[in model A,][]{Ferrers77}, the Dehnen 
\citep[in model D,][]{Dehnen00} and the Barbanis-Woltjer \citep[in
  model BW,][]{Barbanis.Woltjer67} potentials. The two last ones are ad
hoc, but have the big advantage of providing a considerable
non-axisymmetric force beyond corotation. Ferrers bars correspond to a
realistic projected density distribution, but their non-axisymmetric
force beyond corotation decreases strongly with radius and their
vertical mass distribution is oversimplified. It would thus be both
straightforward and useful to extend this work to other 
potentials. 

We need to check how different potentials affect
the various correlations and trends we found and whether these become
substantially broader when results from many different but all
realistic potentials are used. But the number of possible potentials is
very large and we can never be sure that we have covered all the relevant
ones, while some of the potentials may be unrealistic and thus
introduce an artificial broadening. Thus a better possibility,
outlined already in Paper II, is to 
calculate potentials directly from galaxy images and the galaxy
rotation curve, the former preferably in a wavelength tracing the old
stellar population  \citep[e.g.][]{Lindblad.LA96, Salo.RBPCCL99, Kranz.SR03,
  Perez.FF04, Byrd.FB06, Rautiainen.SL08}.
Although this is in principle straightforward, it still has two
considerable drawbacks. The first one is that even in the
near infrared a non-negligible fraction of the light is contributed by
young stars and not from the old underlying stellar population and 
is not easy to eliminate. The second difficulty comes from the fact
that we observe galaxies as projected on
the sky and have no information on their three-dimensional structure,
other than by extrapolating from similar galaxies seen at other
orientations. Two identical projected density distributions can, 
however, have considerably different three dimensional
density distributions and potentials. This introduces a particularly
serious drawback 
in the bar region. Indeed we know that, seen edge-on, bars are 
vertically thin only for a short time after their formation and afterwards
most of their extent acquires a boxy or peanut shape which extends
vertically well above and below the galactic disc \citep[see e.g.][for a
  review]{Atha08}. This can introduce 
non-negligible errors in the calculation of bar potentials and forcings
from near infrared images. Even so, potentials stemming directly from
observations can be most
useful for calculating orbits and manifolds, since it is guaranteed
that, to a first approximation, they correspond to realistic density
distributions.  

There is a further way in which potentials should be improved, namely
by including evolution. Indeed it is well known that barred
galaxies evolve and thus it would be useful to calculate manifolds in
evolving, i.e. time dependent potentials. Such work is underway
(Romero-G\'omez et al, in preparation). 

Even a perfect potential, however, can give only
qualitative information about the orbits that constitute it and can
not reveal whether a given manifold guides a sufficient number of
orbits for a structure to be observable. Such information on the orbits 
can come only from  $N$-body simulations, or from methods
such as the Schwarzschild method and its extensions
\citep[][etc.]{Schwarzschild79, Syer.Tremaine96, deLorenzi.DGS07,
  Dehnen09}, or the iterative scheme \citep{RodionovAS09}. Such methods  
provide initial conditions for orbits, thus making it possible to test
whether the latter are trapped by the manifolds. 

Finally, a most important contribution to assessing our theory will
come from further comparisons with observations and by checking out
the observational predictions made in this series of papers. 

\section{Summary and conclusions}
\label{sec:summary}

Since the early work of B. Lindblad (\citeyear{Lindblad63}), density waves have
been commonly assumed to be at the basis of any explanation of spiral
structure formation. In this series of papers we presented
an alternative viewpoint, which we have applied
specifically to barred galaxies. This explains the formation of spiral
arms, as well as of inner and outer rings, in a common theoretical
framework. According to our theory, it is the unstable Lagrangian points
located at the ends of the bar and the corresponding
manifolds that are responsible for the formation of 
spirals and rings. These manifolds drive orbits, which are in fact
chaotic, but are confined by the manifolds, so that they create
over-densities which have the right shape to explain the spirals and
the rings (Papers I and II). 

We showed that weaker non-axisymmetric perturbations will produce
manifolds of $R_1$ ring shape, while stronger ones will produce other
types of rings and spirals. It is, nevertheless, the same theory that
can explain all these different morphologies. This is in good
agreement with 
the observed morphological continuity between rings, pseudorings and
spirals. This continuity is clear in our theory, but not in others.

We made many different kinds of comparisons with observations. 
We first made sure that our theory can account for the basic
morphological structures observed, namely the spirals, the inner rings
and the outer rings (Paper IV). The model outer rings have the observed $R_1$,
$R_2$ and $R_1R_2$ ring and pseudo-ring size and morphologies,
including the dimples 
often observed near or at the direction of the bar major axis. Both model
and observed inner rings have roughly the same relative sizes and
orientations. 

We next turned to the main spiral arm properties (Paper IV). We showed that, as
the bar grows, material gets trapped mainly in the unstable manifolds,
i.e. the sense of the arms will be trailing, in good agreement with
what is currently known from observations. There is, nevertheless, a
very small fraction of the material concerned that will be caught by
the stable manifold branch, and will thus form a very low amplitude
leading spiral. If its amplitude is not too low, it may be observable
in a Fourier analysis of the galaxy image, or as interference patterns
on the arm amplitude.

Theoretical and observational results agree also well on the number of
arms in a barred galaxy (Paper IV). Our theory links one spiral arm to each of
the unstable Lagrangian points in the standard case. Since galactic
bars are bisymmetric, there should be two such Lagrangian points, the
$L_1$ and $L_2$, and therefore two spiral arms. This morphology
persists even in the non-standard case where the $L_1$ and $L_2$ are
stable (Paper III). Indeed, the vast majority of barred galaxies have
two spiral arms. There are, nevertheless, a few barred galaxies with four
spiral arms. Our theory can account for such cases, given the right
potential, and we discussed a few such potentials in Paper IV. 

We also compared the shape of observed spiral arms with those of
manifolds (Paper IV). We showed that our theory can explain the characteristic
arm winding often observed, namely that, as the angle along the arm
increases, the radius first increases and then, after
reaching a maximum, `falls back' towards the bar. This property has,
to our knowledge, 
not been reproduced by any other theory so far, but follows naturally from
ours. We also predict that spirals in galaxies with stronger bars will be
more open than those in galaxies with weaker bars. This prediction can
be made because our theory, contrary to most others, is non-linear.
Finally, very tightly wound near-logarithmic
spirals can also  be obtained with our theory, but will rely
considerably on spiral forcing.

The shape of rings and the ratio of their extents, although not as
straightforward to compare as other properties, reveal interesting
information (Paper IV). Concerning $R_1$ 
rings, for which we have sufficient models to get adequate statistics,
we find that the ring shapes cover the same range as observations.
A comparison for inner rings is less straightforward, because in
strong bar cases parts of the inner manifold
branches outline parts of the bar and not an inner ring. Taking this
into account, we find that there is a fair agreement between models
and observations.

One of the predictions made in Paper IV concerns the shape of the inner
rings. Namely, we found a strong correlation between the ring shape
and $Q_{t,L_1}$, in the sense that bars with a stronger forcing at or
somewhat beyond $L_1$ should have more elongated $R_1$ rings. We also
stressed that, in order for correlations concerning a spiral or a ring
property with the strength of the non-axisymmetric forcing to be
strong, the latter quantity should be measured at or beyond the
Lagrangian radius, i.e. with $Q_{tL_1}$. If $Q_b$ is used instead, the
correlation can be destroyed, or severely damaged. Both of these results have
now been confirmed by observations \citep{GrouchyBSL10}. 

We also introduced collisions and dissipation to the manifold
calculations, in order to roughly model the gas properties (Paper IV).
We found that this does not
influence the existence of the spiral arms or rings, and makes no
major modifications to their
shape. The amount of dissipation, however, does influence the width of
the arms. These become thinner as the dissipation is increased, so that
the gaseous arm comes nearer to the lowest energy manifold. 

Our theory makes very clear and, for many morphological types, precise
predictions about the value of 
the bar pattern speed, since it shows how it is possible from the
morphology only to locate the positions of the $L_1$ and $L_2$, 
both for spirals and for the various ring morphologies (Paper V). This of
course works only in the case where sufficient morphological features
are present in the observed galaxy. For example if there is only a 
bar, with no spirals or rings, our theory can not be applied and the
$L_1$ and $L_2$ can not be located. However, in cases with a reasonable
amount of structure our theory allows the location of the $L_1$ and
$L_2$ from morphology alone. In cases with appropriate morphological
features this may allow a more accurate determination of the pattern
speed then other methods used so far. 
Our theory also provides building blocks to explain the
rectangular-like outline observed in strong barred galaxies and to
explain ansae (Paper V). 

Concerning photometry, our results are not very constraining. The
amplitude of the spiral should in general decrease outwards, but this
decrease may well not be monotonic, because other arm components
-- e.g. a weak, leading component --, if present, 
will lead to bumps on an otherwise smoothly decreasing density profile. 

The circulation of material within the manifolds also depends on the
bar strength. In
the relatively weak bars, which form $rR_1$ morphologies, the mass
elements can move from the inner manifolds to the outer manifolds and then
back to inner ones again. I.e. material can move from the region
within corotation to the region outside it and vice versa via the
neighbourhood of the $L_1$ and $L_2$. Averaged over time, this brings
internal circulation, but no net motion of material inwards or
outwards. The situation is more complex for the cases where the
non-axisymmetric forcing around corotation is stronger and which have
a spiral morphology. In such cases, 
material moves from the region within corotation to the region outside
it. If the potential has a predominantly barred structure, this material
returns eventually to the region of the bar or of the other arm.
But if the potential has also a spiral component, then the arm can
continue winding outwards for much longer times. Either way, this
results in a considerable radial mixing and material can reach
the outermost parts of the disc and can
contribute to its radial extension. Of course the density wave
theory had already predicted that in a spiral galaxy the radial motion
is outwards outside corotation \citep{Lin.Shu71,
  Kalnajs78}. Our results, however, are more than a simple 
restatement of the above, because our theory shows how material from well
within corotation can reach regions well outside it. It can thus
explain considerable radial mixing and have a
more important impact on the formation of the outer disc than those of
the density wave theory. Furthermore, our theory predicts the
relation between bar strength and the abundance gradients found in
observations \citep{Martin.Roy94}. 

We also made predictions on the galaxy kinematics, by measuring
the radial and tangential velocity components along the manifold
loci, i.e. along the spirals and the inner and outer rings. For the
tangential component we subtract the local circular velocity and thus
obtain the perturbation, or peculiar tangential velocity. Plotted as a
function of the azimuthal angle along the ring, both radial and tangential
velocities show $m$ = 2 oscillations, the amplitude of the former being
higher than that of the latter. We find that the
amplitude of these oscillations increases with increasing bar strength
and that they are larger in the inner than in the outer ring. The
latter result is in agreement with the former, since the bar forcing
is higher at the inner ring than at the outer one. 

For spirals, we find that the radial
velocity of material moving from the Lagrangian points outwards
along the arm increases with the distance to the Lagrangian point
until it reaches a maximum and then it decreases, until at some point
it changes sign. From that point onwards the arm will start
approaching the other arm or the region of the bar. Where this occurs
is a function of the bar strength and, when observed, will provide
an additional confirmation to our theory. Nevertheless, this may be a
difficult observation to make, since the spiral density is low in
that region. Furthermore, in galaxies with a high spiral amplitude, the spiral
forcing may be a considerable fraction of the non-axisymmetric
forcing, so that this sign reversal may not take place.
Finally, we calculated the
line-of-sight velocity and found that in the case of rings its
variation along the ring can be well modelled by a sinusoidal. This is
not true for very strong bars and spiral morphologies. Finally, when the
sinusoidal is a reasonable fit, we find that the difference between
the kinematic and photometric major axes is in 
agreement with the observed values.

As already mentioned, our models show a clear connection between bar
strength and 
morphology of the response. Relatively weak non-axisymmetric forcings
give $R_1$ and $R_1'$ morphologies, while stronger forcings give
spirals and other types of rings. Yet it is possible in 
self-consistent calculations, in which the potential due to the material
trapped in the manifold is taken into account, to form a spiral in
cases where the bar on its own is too weak to sustain it. This is due
to the fact that initially the only part of the ring that will be
populated is the part which is near the $L_1$ or the $L_2$. This then
adds a spiral-like forcing to that of the bar, so that manifold will
have a more spiral-like shape. It is thus possible in self-consistent
calculations to have a spiral morphology in a bar potential which
is somewhat below the threshold between the $R_1$ and the spiral
morphology in the rigid bar forcing case. 

In these series of papers, we applied our theory to barred
galaxies. It can, however, be straightforwardly applied also to
non-barred galaxies with 
other non-axisymmetric perturbations, such as spirals or ovals,
provided these have unstable saddle Lagrangian points. There is thus
no discontinuity between barred and non-barred spiral galaxies. On the
other hand it is unclear at this point whether, and under what
conditions, complex 
unstable Lagrangian points can have the type of manifolds that can
create realistic spirals or rings. Such Lagrangian points are found
e.g. at the $L_4$ and $L_5$ Lagrangian points of very strong bars
\citep[][and Paper IV]{Atha92a}. More work is necessary before this
question can be answered. 

We discussed how material can get trapped in the manifolds, where the
mass in the spirals comes from and whether the spirals are short-lived
or long-lived. We also compared our theory with other theories not
relying on chaotic orbits and manifolds. In particular, we discussed a
number of advantages which our theory has over the density wave
theory. We stressed that it need not
necessarily be the same theory that explains all spirals in disc
galaxies, and even in one single galaxy more than one mechanism can be at
play. The properties of the resulting spiral would then result from
the self-consistent interaction of all the spirals of different
origin. We finally propose a number of avenues for future work within
the framework of our theory.

We showed that the orbits that form the building blocks of the spirals
and of the inner and outer rings are chaotic. Nevertheless, they can
form narrow features because they are confined by the manifolds, which
act as guiding tubes. We propose to call this type of chaos `confined
chaos' and suggest that such orbits should be taken into account in
dynamical studies, since they could contribute, together with the
regular orbits trapped around the stable periodic orbits, to galactic
structures. 

A prerequisite for any theory is that it can be falsified
\citep[e.g.][]{Popper59}. Ours fulfils well this criterion. Indeed
there are a number of observational results that 
could have proved it wrong. Namely, if the two armed global spirals
were not preponderant in symmetric barred galaxies, or if these arms
were leading, or if the major features of the observed and the
theoretical morphologies were inconsistent, our theory would have been
proven wrong. As we showed in this paper, this is not the case and
there is good agreement between our theory and observations in these
and many other points. There are more observations which could have
proven our theory to be wrong. For example, our theory would be
invalidated if it was clearly shown that all spirals rotate with a
different pattern speed than the bar.   

In total over the five papers, we presented a number of predictions of our
theory. Whenever the necessary data are available, we compared the
results of our theory to the existing observational results. All tests
we have been able to make tend to show that our theory is viable. In
many cases, however, new observations, or new analysis of existing
data are necessary. In such cases, we presented here simply the
theoretical predictions  
and outlined the way the comparisons could be made. The results of
these comparisons will be very important in order to confirm our
theory, to bring about modifications, or to reject it. As already
mentioned, one of the theoretical predictions of paper IV concerning
the dependence of the inner ring shape on bar strength has
already been confirmed by observations \citep{GrouchyBSL10}.

We can thus conclude that the theory we presented in this series of papers 
should be useful in order to explain the formation of spirals and of
inner and outer rings in barred galaxies. Concerning rings, no other
theory has been fully developed, while our results agree very well with
those of simulations, suggesting that the motion of particles
in them are guided by the manifolds we describe here. Concerning
spirals, the comparisons we have made
with observations are more extensive than those that have been made so
far for the swing amplifier theory. Furthermore, our theory is
clearly falsifiable, contrary to the classical WKBJ density wave theory
(see discussion by \cite{Kalnajs78}). We
thus believe that it should be considered as a possible alternative,
on a par with other theories. More work, both theoretical and
observational, is necessary in order to establish in what cases each one
of the so far proposed theories prevails and how they can, perhaps all
together, come to  an explanation of spiral formation and spiral properties.

\section*{Acknowledgements}

EA thanks Scott Tremaine for a stimulating discussion on the manifold
properties. We also thank Ron Buta, A. Mel'nik and P. Rautiainen for
very useful discussions and email exchanges. This work was partly
supported by grant ANR-06-BLAN-0172 and by the MICINN-FEDER grant 
MTM2009-06973, CUR-DIUE grant 2009SGR859 and  MICINN grant 
AYA2009-14648-C002-01.

\bibliography{manifIIb-after-ref}

\label{lastpage}

\end{document}